\documentclass[twocolumn]{aastex631}
\usepackage{amsmath}

\newcommand{\COiso}{$^{13}\textrm{CO}$}
\newcommand{\COmain}{$^{12}\textrm{CO}$}
\newcommand{\HTWOO}{H$_{2}\textrm{O}$}
\newcommand{\COTWO}{CO$_{2}$}
\newcommand{\methane}{CH$_{4}$}
\newcommand{\ammonia}{NH$_{3}$}
\newcommand{\enstatite}{MgSiO$_{3}$}

\newcommand{\target}{HIP 99770 b}
\newcommand{\primary}{HIP 99770}
\newcommand{\Mjup}{$M_\mathrm{Jup}$}
\newcommand{\Rjup}{$R_\mathrm{Jup}$}
\newcommand{\Msun}{$M_\odot$}
\newcommand{\Teff}{$T_\mathrm{eff}$}
\newcommand{\kms}{km\ s$^{-1}$}

\newcommand{\caltech}{Department of Astronomy, California Institute of Technology, Pasadena, CA 91125, USA}
\newcommand{\gps}{Division of Geological \& Planetary Sciences, California Institute of Technology, Pasadena, CA 91125, USA}
\newcommand{\ucsc}{Department of Astronomy \& Astrophysics, University of California, Santa Cruz, CA95064, USA}
\newcommand{\keck}{W. M. Keck Observatory, 65-1120 Mamalahoa Hwy, Kamuela, HI, USA}
\newcommand{\ucla}{Department of Physics \& Astronomy, 430 Portola Plaza, University of California, Los Angeles, CA 90095, USA}
\newcommand{\jpl}{Jet Propulsion Laboratory, California Institute of Technology, 4800 Oak Grove Dr.,Pasadena, CA 91109, USA}
\newcommand{\ucsd}{Department of Astronomy \& Astrophysics,  University of California, San Diego, La Jolla, CA 92093, USA}

\newcommand{\osu}{Department of Astronomy, The Ohio State University, 100 W 18th Ave, Columbus, OH 43210 USA}

\newcommand{\arizona}{James C. Wyant College of Optical Sciences, University of Arizona, Meinel Building 1630 E. University Blvd., Tucson, AZ. 85721}

\shorttitle{HIP 99770 b with KPIC}
\shortauthors{Zhang et al.}

\graphicspath{{./}{figures/}}

\begin{document}

\title{Atmospheric characterization of the super-Jupiter HIP 99770 b with KPIC}

\author[0000-0003-0097-4414]{Yapeng Zhang}
\altaffiliation{51 Pegasi b fellow}
\affiliation{\caltech}

\author[0000-0002-6618-1137]{Jerry W. Xuan}
\affiliation{\caltech}

\author{Dimitri Mawet}
\affiliation{\caltech}

\author[0000-0003-0774-6502]{Jason J. Wang}
\affiliation{Center for Interdisciplinary Exploration and Research in Astrophysics (CIERA) and Department of Physics and Astronomy, Northwestern University, Evanston, IL 60208, USA}

\author[0000-0002-5370-7494]{Chih-Chun Hsu}
\affiliation{Center for Interdisciplinary Exploration and Research in Astrophysics (CIERA) and Department of Physics and Astronomy, Northwestern University, Evanston, IL 60208, USA}

\author[0000-0003-2233-4821]{Jean-Bapiste Ruffio}
\affiliation{\ucsd}

\author[0000-0002-5375-4725]{Heather A. Knutson}
\affiliation{\gps}

\author[0000-0001-9164-7966]{Julie Inglis}
\affiliation{\gps}

\author{Geoffrey A. Blake}
\affiliation{\gps}

\author[0000-0003-1728-8269]{Yayaati Chachan}
\altaffiliation{CITA National Fellow}
\affiliation{Department of Physics and Trottier Space Institute, McGill University, 3600 rue University, H3A 2T8 Montreal QC, Canada}
\affiliation{Trottier Institute for Research on Exoplanets (iREx), Universit\'e de Montr\'eal, Canada}

\author[0000-0001-9708-8667]{Katelyn Horstman}
\affiliation{\caltech}

\author{Ashley Baker}
\affiliation{\caltech}

\author{Randall Bartos}
\affiliation{\jpl}

\author[0000-0003-4737-5486]{Benjamin Calvin}
\affiliation{\ucla}

\author{Sylvain Cetre}
\affiliation{\keck}

\author[0000-0001-8953-1008]{Jacques-Robert Delorme}
\affiliation{\keck}

\author{Greg Doppmann}
\affiliation{\keck}

\author[0000-0002-1583-2040]{Daniel Echeverri}
\affiliation{\caltech}

\author[0000-0002-1392-0768]{Luke Finnerty}
\affiliation{\ucla}

\author[0000-0002-0176-8973]{Michael P. Fitzgerald}
\affiliation{\ucla}

\author[0000-0001-5213-6207]{Nemanja Jovanovic}
\affiliation{\caltech}

\author[0000-0002-4934-3042]{Joshua Liberman}
\affiliation{\caltech}
\affiliation{\arizona}

\author[0000-0002-2019-4995]{Ronald A. L\'opez}
\affiliation{\ucla}


\author{Evan Morris}
\affiliation{\ucsc}

\author{Jacklyn Pezzato}
\affiliation{\caltech}


\author[0000-0003-1399-3593]{Ben Sappey}
\affiliation{\ucsd}

\author{Tobias Schofield}
\affiliation{\caltech}

\author{Andrew Skemer}
\affiliation{\ucsc}

\author[0000-0001-5299-6899]{J. Kent Wallace}
\affiliation{\jpl}

\author[0000-0002-4361-8885]{Ji Wang}
\affiliation{\osu}

\author[0000-0001-5173-2947]{Clarissa R. Do \'O}
\affiliation{Department of Physics, University of California, San Diego, La Jolla, CA 92093, USA}

\begin{abstract}

Young, self-luminous super-Jovian companions discovered by direct imaging provide a challenging test of planet formation and evolution theories. By spectroscopically characterizing the atmospheric compositions of these super-Jupiters, we can constrain their formation histories.
Here we present studies of the recently discovered \target, a 16 \Mjup~high-contrast companion on a 17 au orbit, 
using the fiber-fed high-resolution spectrograph KPIC ($\mathcal{R}\sim 35,000$) on the Keck II telescope. 
Our K-band observations led to detections of \HTWOO\ and CO in the atmosphere of \target.
We carried out free retrieval analyses using \texttt{petitRADTRANS} to measure its chemical abundances, including the metallicity
and C/O ratio,  projected rotation velocity ($v\sin i$), and radial velocity (RV). 
We found that the companion's atmosphere has C/O $=0.55_{-0.04}^{+0.06}$ and [M/H] $=0.26_{-0.23}^{+0.24}$ (1$\sigma$ confidence intervals), values 
consistent with those of the Sun and with a companion formation via gravitational instability or core accretion. 
The projected rotation velocity
$v\sin(i) < 7.8$ \kms\ is small relative to other directly imaged companions with similar masses and ages. 
This may imply a near pole-on orientation or effective magnetic braking by a circumplanetary disk.
In addition, we added the companion-to-primary relative RV measurement to the orbital fitting and 
obtained updated constraints on orbital parameters.
Detailed characterization of super-Jovian companions within 20 au like \target\ is critical for understanding 
the formation histories of this population.

\end{abstract}


\section{Introduction} \label{sec:intro}

Young, self-luminous super-Jovian companions discovered by direct imaging surveys have posed 
significant challenges to planet formation theories. 
These objects typically straddle the mass boundary between planets and brown dwarfs ($\sim$13 \Mjup) and 
orbit their primary stars at large separations ($>$10 au). 
Their formation mechanisms remain under debate as they are not easily compatible with either
in-situ core accretion or gravitational instability \citep{PollackEtAl1996, Boss1997, Chabrier2003, KratterLodato2016}. 
The long timescale of core accretion in the outer protoplanetary disk makes it challenging to assemble  
massive cores to trigger runaway accretion before the gas disk dissipates \citep{LambrechtsJohansen2012}, while
disk instability and cloud fragmentation are expected to form more massive objects 
like brown dwarfs or stellar companions \citep{ZhuEtAl2012, ForganRice2013a, OffnerEtAl2023}.
For some wide-orbit companions at hundreds of au, 
formation at closer orbits followed by scattering events may be effective \citep{VerasEtAl2009}.
Previous studies on orbital architectures of these super-Jupiters ($<20$ \Mjup) versus those of brown dwarfs ($>20$ \Mjup)
found evidence for distinct distributions in their semimajor axes and orbital eccentricities \citep{NielsenEtAl2019, BowlerEtAl2020, NagpalEtAl2023, DoOEtAl2023}. 
This suggests that the two populations likely have distinct formation mechanisms, but it is unclear
whether there is a well-defined boundary separating them.

Atmospheric characterization provides an important avenue to distinguish between competing formation mechanisms for substellar companions.
Elemental abundances including the metallicty [M/H] and carbon-to-oxygen (C/O) ratios 
have been suggested as probes for planet formation \citep[e.g. ][]{ObergEtAl2011, Madhusudhan2012, TurriniEtAl2021}.
In general, formation via gravitational instability or cloud fragmentation is thought to result in compositions akin to stars,
while core accretion can result in a variety of compositions depending on the 
accretion mechanisms, formation location relative to disk icelines, migration histories 
\citep[e.g.][]{MadhusudhanEtAl2014, MordasiniEtAl2016, BitschEtAl2019}, among other factors. 
Although companions formed via gravitational collapse can also be enriched through late accretion, the impact on their compositions is not expected to be significant because of the high envelope masses \citep{InglisEtAl2024}.

Super-Jupiters on wide orbits are excellent targets for directly probing planetary emission at high signal-to-noise, 
therefore allowing for robust constraints on their chemical compositions.
As measurements of atmospheric C/O and metallicity in directly imaged super-Jupiters have accumulated over the past few years,
we begin to build a sample large enough to test the hypotheses.
Spectral surveys of substellar companions with masses of $10-30$ \Mjup~such as \cite{HochEtAl2023, XuanEtAl2024b} suggest that they display broadly solar compositions, which are chemically compatible with
formation via gravitational instability or core accretion beyond CO icelines, where solids have approximately stellar C/O ratios.
Although the planetesimal accretion scenario is less likely for wide-orbit massive gas giants, pebble accretion, whose efficiency increases in the outer disk, is not precluded \citep{LambrechtsJohansen2012, BitschEtAl2015, BitschEtAl2019}.

At an individual system level, linking the atmospheric composition of a planet to its formation history 
is complicated by a chain of processes, including dust and ice/vapor chemistry evolution in disks, 
planet migration, late solid enrichment, and mixing of planet's core and envelope \citep{MolliereEtAl2022a}. These convoluted processes make inferring a planet's formation history from its current atmosphere difficult.
The challenging nature of this task calls for combining different lines of evidence 
from atmospheric tracers such as refractory-to-volatile ratios  \citep{LothringerEtAl2021, ChachanEtAl2023}
and isotopologue ratios  \citep{ZhangEtAl2021, ZhangEtAl2021a} as well as dynamical indicators 
such as spin and orbital architecture \citep{BryanEtAl2018,  BryanEtAl2020a, xuan_Rotation_2020}.
High-resolution spectroscopy of super-Jovian companions provides information on these various aspects simultaneously, 
enabling a more comprehensive understanding of the current state of planetary system and its potential connection to  
formation pathways and evolutionary history.

Despite extensive studies of wide-orbit companions \citep[e.g.,][]{ZhangEtAl2021a, Palma-BifaniEtAl2023, Petrus2024, InglisEtAl2024, XuanEtAl2024b}, there are relatively few published spectral characterization of companions with orbital separation within 20 au, 
which represents a critical region of the parameter space for understanding 
population-level trends.
Most super-Jovian companions with compositional measurements are located beyond 50 au, 
with only a handful of exceptions such as $\beta$ Pic b \citep{GRAVITYCollaborationEtAl2020, LandmanEtAl2024}, 
HR 8799 cde \citep{KonopackyEtAl2013, MolliereEtAl2020, RuffioEtAl2021, WangEtAl2022, NasedkinEtAl2024}, 
51 Eri b \citep{Brown-SevillaEtAl2023,WhitefordEtAl2023}, 
and AF Lep b \citep{ZhangEtAl2023, Palma-BifaniEtAl2024}. 
Planets with cooler temperatures and/or smaller separations are more challenging 
to characterize because of the predominant stellar speckle contamination. 
Techniques that combine both the spatial and spectral resolving capability 
have proven to be powerful for enhancing the contrast limit of detection and characterization 
\citep{SnellenEtAl2015, HoeijmakersEtAl2018, RuffioEtAl2023, Agrawal2023}.
New instruments coupling adaptive optics systems with high-resolution 
spectrographs, such as the Keck Planet Imager and Characterizer \citep[KPIC,][]{MawetEtAl2017}, 
HiRISE on VLT \citep{ViganEtAl2023}, and REACH on Subaru \citep{KotaniEtAl2020a}, have recently enabled high resolution spectroscopy of faint companions on close-in orbits \citep[e.g., HR 8799 planets;][]{WangEtAl2021}.

The recently discovered 16 \Mjup~\target~is among only a handful of super-Jovian companions 
with an orbital separation within 20 au \citep{CurrieEtAl2023}. Its current location is expected to be within the
CO iceline of the protoplanetary disk around A-type host stars \citep{QiEtAl2015}.
In core-accretion scenarios, the C/O ratio likely varies as a function of companion's metallicity because of the partitioning of solid versus gas compositions.
Therefore, companions at smaller separations are
likely to display a larger variety of atmospheric compositions,
providing intriguing case studies for testing formation models by measuring both the C/O and metallicity.
In contrast, beyond the CO iceline, the companion's C/O ratio is expected to be solar for a wide range of metallicity because most volatiles are condensed out.
We note, however, if the companion has a solar metallicity, it is difficult to distinguish between core accretion and gravitational instability as both mechanisms can plausibly lead to this outcome, regardless of the planet's birth location.

In this paper, we present the atmospheric characterization of \target~
using KPIC observations.
When combined with its spin velocity and orbital properties, this allows us to
place constraints on its formation and evolution.
The paper is organized as follows. In Section~\ref{sec:system}, we introduce the target system
HIP 99770. The KPIC observations and data reduction are explained in Section~\ref{sec:observation}.
Then, we carry out free retrieval analyses with the observations. 
The forward model and retrieval framework are described in Section~\ref{sec:retrieval}.
We present the molecular detections and retrieval results in Section~\ref{sec:result}.
We discuss the measured chemical abundances, spin velocity, 
and their implications for planet formation in Section~\ref{sec:discussion}.
Finally, we summarize the findings in Section~\ref{sec:conclusion}.

\section{HIP 99770 system} \label{sec:system}

\target~is a recently discovered super-Jovian companion with a joint direct imaging and precision astrometry 
detection \citep{CurrieEtAl2023}. 
We summarize the properties of the \primary~system in Table~\ref{tab:system}.
The 1.8 \Msun~A-type primary star \primary~has an effective temperature of $\sim$8000 K, an age of 40-400 Myr, 
and a distance of 40.74 pc 
\citep{GaiaCollaborationEtAl2021, CurrieEtAl2023}.
The primary star shows evidence of 
Hipparcos-Gaia astrometric acceleration induced by an orbiting companion, 
which was confirmed 
by direct imaging with the Subaru Coronagraphic Extreme Adaptive Optics Project 
(SCExAO/CHARIS) and Keck/NIRC2 at an angular separation of 0.44\arcsec~\citep{Brandt2021, CurrieEtAl2023}. 
The combination of the relative astrometry and
acceleration leads to a dynamical mass of $16.1^{+5.4}_{-5.0}$ \Mjup\ for the companion, 
or a companion-to-star mass ratio of $\sim 8\times10^{-3}$, at a semimajor axis of 
$16.9^{+3.4}_{-1.9}$ au and an orbital eccentricity of $0.25^{+0.14}_{-0.16}$  
\citep{CurrieEtAl2023}. 
The companion's luminosity and dynamical mass are consistent with an age of 115-200 Myr. 

\cite{CurrieEtAl2023} carried out spectral characterization of \target~
using a CHARIS JHK broadband spectrum at 1.16-2.37 $\mu$m and Keck/NIRC2 L band photometry, 
which suggested a spectral type of L7-L9.5 near the L/T transition, 
an effective temperature of \Teff $\sim$ 1300-1600 K, and a surface gravity 
of $\log g =$4-4.5. Evolutionary models \citep{BaraffeEtAl2003, SpiegelBurrows2012} predict a radius of 1.1-1.2 \Rjup~while the 
spectral analysis tends to imply a smaller radius of up to 1.05 \Rjup~\citep{CurrieEtAl2023}.
A comparison to HR 8799 d's spectral shape shows that \target's spectrum is 
less flat, indicating moderate cloudiness.

\begin{deluxetable}{lcc}
    \tablecaption{\label{tab:system}Properties of the \primary~system}
    \tablehead{\colhead{Property} & \colhead{Value} & \colhead{References}}
    \startdata
    \multicolumn{3}{l}{\primary} \\
    \hline
    $\alpha_{2000.0}$ & 20:14:32.032 & 1 \\
    $\delta_{2000.0}$ & +36:48:22.7 & 1 \\
    Distance (pc) & $ 40.74 \pm 0.15 $ & 1 \\
    $v_\mathrm{sys}$ $\textrm{(\kms)}$   & $-20.52 \pm 0.40$ & 1 \\
    $v\sin i$ $\textrm{(\kms)}$ & $81.1 \pm 1.5$ & 1 \\
    Age (Myr) & 40 (Argus membership)  & 2 \\
     & 115-414 (astroseismology) & 2 \\
    Mass ($M_\odot$) & $ 1.85 \pm 0.19$ & 2 \\
    SpT & A5-A6 & 3 \\
    \Teff (K) &  8000  &  3 \\
    $K_s$ (mag) & $4.42 \pm 0.02$ & 2 \\
    $L_p$ (mag) & $4.40 \pm 0.05$ & 2 \\
    $\rm [C/H]$ & $0.18 \pm 0.09$ & 4 \\
    $\rm [O/H]$ & $0.01 \pm 0.09$ & 4 \\
    \hline
    \multicolumn{3}{l}{\target} \\
    \hline
    SpT & L7-L9.5 & 2\\
    \Teff (K) &  $1400^{+200}_{-150} $ &  2\\
    $K_s$ (mag) & $15.66 \pm 0.09$ & 2\\
    $L_p$ (mag) & $14.52 \pm 0.12$ & 2\\
    $a$ (au) & $16.9^{+3.4}_{-1.9}$ & 2 \\
    eccentricity & $0.25^{+0.14}_{-0.16}$ & 2 \\
    Mass (\Mjup) & $16.1^{+5.4}_{-5.0}$ & 2 \\
    Radius (\Rjup) & $1.0- 1.1$ & 2\\
    $\log g$ (cgs) & $4.0-4.5$ &  2\\
    $v\sin i$ $\textrm{(\kms)}$ & $3.5_{-1.9}^{+1.7}$ & 5\\
    $\rm [M/H]$ &  $0.26_{-0.23}^{+0.24}$ &   5\\
    C/O &  $0.55_{-0.04}^{+0.06}$ &  5 \\
    \hline
    \enddata
    \tablerefs{(1) \citet{GaiaCollaborationEtAl2021}, 
    (2) \cite{CurrieEtAl2023}, (3) \cite{MurphyPaunzen2017}, (4) \cite{ErspamerNorth2003, HinkelEtAl2014},
    (5) this work.}
\end{deluxetable}

\section{Observations and Data Reduction} \label{sec:observation}

We observed the 
system with KPIC \citep{MawetEtAl2017, 
DelormeEtAl2021, EcheverriEtAl2022} on UT 2023 June 16 and 21. 
KPIC uses a fiber injection unit (FIU) located downstream of the Keck II adaptive optics system 
to inject light into single-mode fibers, which are connected to the NIRSPEC high-resolution ($\mathcal{R}\sim 35,000$) spectrograph
\citep{McLeanEtAl1998, MartinEtAl2018} to disperse light onto the detector. 
Among the four illuminated fibers, one of them is aligned 
to the position of the companion to obtain its spectra. 
The use of single-mode fibers aids in the suppression of the stellar speckle and 
sky background, and ensure a stable Gaussian-like line spread function 
(LSF) during observations. 
A more detailed description of the instrument can be found in \cite{DelormeEtAl2021, WangEtAl2021}.

\begin{deluxetable}{ccccccc}
    \tablecaption{KPIC Observations of \target}
    \tablehead{\colhead{Date} & \colhead{$t_\mathrm{int}$ (min)} & \colhead{Airmass} & \colhead{Throughput} & \colhead{S/N}
    }
    \startdata
        2023-06-16  & 90 & 1.1 & $3.4\%$ & 1.0 \\
        2023-06-21  & 60 & 1.1 & $3.3\%$ & 0.8 \\
    \enddata
    \label{tab:obs}
    \tablecomments{The S/N per wavelength channel for the companion is estimated from the residuals (see Fig.~\ref{fig:spec}) that include both the background thermal noise and the photon noise due to stellar speckles.}
\end{deluxetable}

Our KPIC observations of the \target~are summarized in Table~\ref{tab:obs}.
We used an AB nodding scheme that alternately aligns the companion on one of two fibers to facilitate sky/background subtraction.  Eighteen frames of 300s exposure time  were recorded
on June 16 and 11 exposures of 600s on June 21. The total integration time on \target~was 90 minutes and 60 minutes, respectively.
We also switched to the primary star once every hour for modeling telluric absorption and instrument response. 
Each night we took short observations on an M giant star, HIP 95771 and HIP 81497, respectively, which display deep and narrow spectral lines, therefore convenient for the calibration of wavelength solution. 

\begin{figure*}[t!]
    \includegraphics[width=\linewidth]{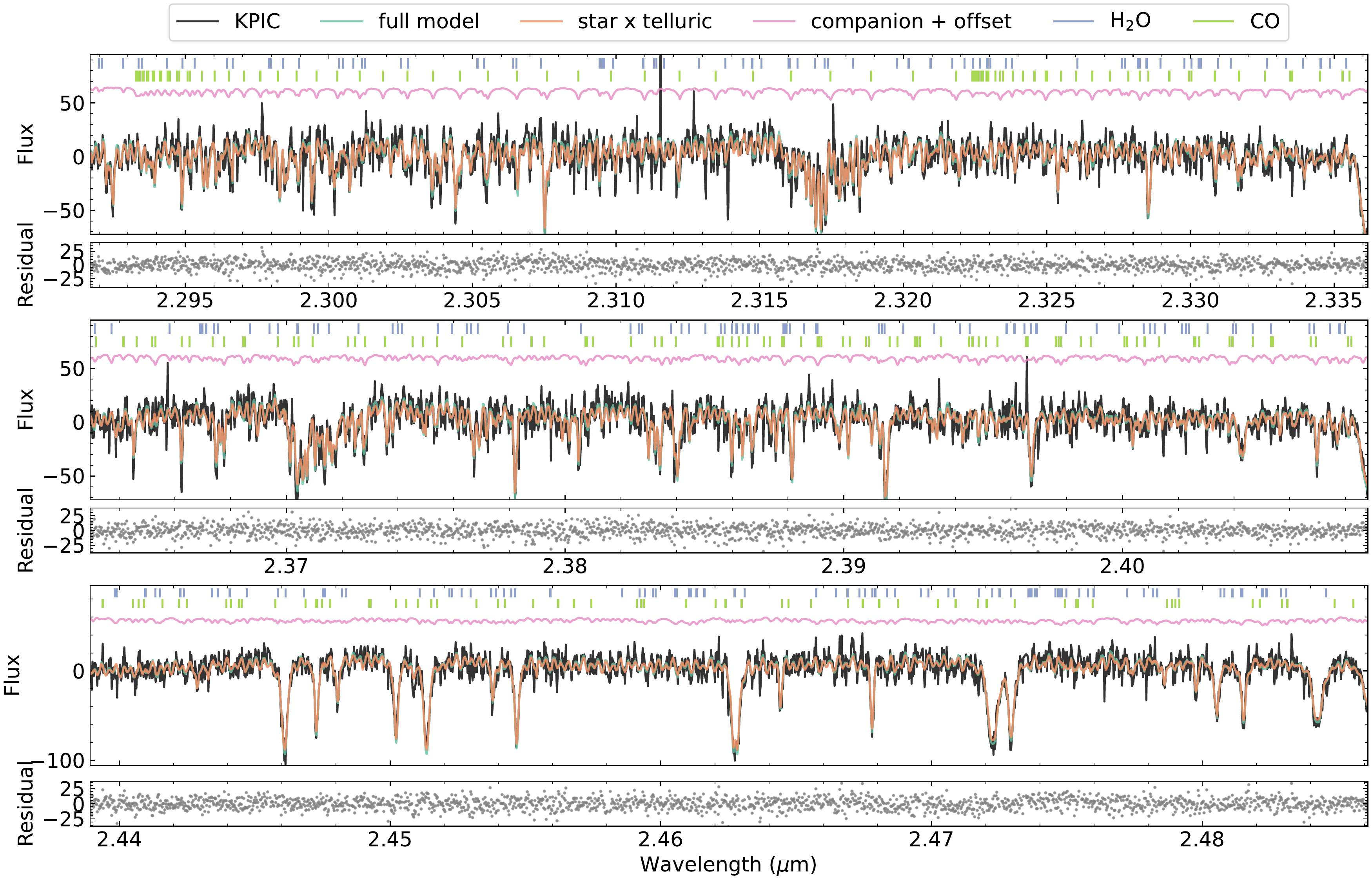}
    \caption{High-pass filtered K-band (2.29-2.49 \micron) 
    spectra of the \primary~system taken with one of KPIC fibers. 
    The observations are shown in black lines. Overplotted are the best-fit full model 
    as obtained with the retrieval analysis.
    The spectrum is dominated by the stellar contribution with a small fraction of emission from the companion, 
    as shown in the offset pink lines. We annotated the positions of absorption features 
    from \HTWOO~and CO with short bars on the top of each panel.
    The residuals (data minus the full model) are shown with scattered dots in gray. 
    \label{fig:spec}}
\end{figure*}

We reduced the data using the public KPIC Data Reduction Pipeline\footnote{\url{https://github.com/kpicteam/kpic_pipeline}}, including steps of 
nod-subtraction, bad pixel removal, spectral order tracing, optimal spectrum extraction, 
and wavelength calibration. We refer readers to \cite{WangEtAl2021} for
an in-depth description. 
The K-band observations span the wavelength range of 1.9-2.5 $\mu$m with a spectral 
resolution of $\mathcal{R}\sim$35,000. We focused our analysis on the three reddest 
spectral orders from 2.29 to 2.49 $\mu$m, covering the main near-infrared absorbing molecules  
such as CO and \HTWOO~in the companion's atmosphere, as shown in Fig.~\ref{fig:spec}. 

As a result of the companion's small separation (0.44\arcsec) and high contrast 
($\Delta K_\mathrm{mag}=11.3$), the extracted spectra are dominated by the 
diffracted starlight, which varies across wavelengths.
Therefore, the continuum of the companion's spectrum cannot be recovered. 
Taking into account the photon noise from the stellar speckles, we reach a signal-to-noise 
(S/N) of $\sim$1 per pixel for the companion's signal in 1.5-hour integration on the first night and S/N $\sim$0.8 on the second night.

\section{Retrieval Analysis} \label{sec:retrieval}

\subsection{Spectral Model of \target} \label{sec:forward}

The spectral model of the companion consists of three components, including the temperature profile, the chemistry model, and the cloud model. The models are set up as follows. 

\subsubsection{Temperature Model}

We parametrize the temperature-pressure (T-P) profile using the gradients 
$\Delta \ln T/ \Delta \ln P$ across 7 pressure knots 
set at $10^{-5}$, $10^{-3.5}$, $10^{-2}$, $10^{-1}$, $10^{0}$, $10^{1}$, 
and $10^{2}$ bar,
and one absolute temperature value, $T_0$, at 100 bar.
This follows the approach in \cite{ZhangEtAl2023, XuanEtAl2024a}.
We set broad prior ranges 
with a uniform distribution to ensure the flexibility of the temperature model (see Table~\ref{tab:params}). 
The full T-P profile is determined by a spline interpolation 
onto 60 layers evenly spaced in log pressure between $10^{-5}$ and 100 bar.

\subsubsection{Chemistry Model}

We explore two different chemistry models: one a free retrieval and the other a disequilibrium model.
The disequilibrium chemistry model first assumes chemical equilibrium to determine 
the chemical abundances at a given pressure, temperature, carbon-to-oxygen ratio (C/O) 
and metallicity ([M/H]), by interpolation of a precomputed table \citep{MolliereEtAl2017, MolliereEtAl2020}. 
The model then parameterizes the disequilibrium chemistry of major species (CO, \methane, and \HTWOO) by enforcing constant 
volume mixing ratios (VMRs) at altitudes above a certain quenching pressure level ($P_\mathrm{quench}$), 
which approximates the effect of rapid vertical mixing in disequilibrium chemistry \citep{ZahnleMarley2014}.
The free chemistry model allows the abundance of each chemical species to vary while assuming a vertically constant profile.
In the case of chemical quenching at a deep atmospheric region below the photosphere, and modest vertical sampling of the atmosphere, 
this constant VMR profile can be a reasonable approximation, which is therefore expected to lead to similar retrieval results as the disequilibrium chemistry model.

\subsubsection{Cloud Model}

We adopt the condensate cloud model from \citet{AckermanMarley2001} 
with \enstatite~as the cloud opacity source -- the expected dominant cloud 
species in L dwarfs \citep{CushingEtAl2006}. 
The cloud is characterized by four parameters: the mass fraction 
of the cloud species at the cloud base $X_0^\mathrm{MgSiO_3}$, the settling parameter $f_\mathrm{sed}$ 
(controlling the thickness of the cloud above the cloud base), 
the vertical eddy diffusion coefficient $K_\mathrm{zz}$ (effectively determining the particle size), 
and the width of the log-normal particle size distribution $\sigma_g$.
Following \cite{MolliereEtAl2020}, the location of the cloud base $P_\mathrm{base}$ is 
determined by intersecting the condensation curve of the cloud species with the T-P 
profile of the atmosphere.

\subsection{Fringing removal}

Inspecting the observational residuals, i.e., the observed spectra minus the best-fit model as obtained through the 
retrieval analysis, we identified sinusoidal fringing features in the data. 
These fringing signals are believed to result from the entrance window of NIRSPEC and KPIC dichroics.
Although the fringing due to the NIRSPEC entrance window is static, 
the KPIC fringing signal shows temporal variation depending on the angle of incidence into the optics and 
the change of optical properties of the material with ambient temperature \citep{FinnertyEtAl2022}. 
We found significant fringing in the periodogram of the residual data 
with a period of $\sim4.5$ \AA\ originating from the KPIC dichroics, especially in the second epoch. 
To model the fringing, we used the function as described in
\cite{XuanEtAl2024a}:
\begin{equation}
    t = \bigg[1+ F\sin^2\bigg( \frac{2\pi n d}{\lambda} \bigg)\bigg]^{-1},
\end{equation}
where $n$ is the index of refraction of the material as a function of wavelength $\lambda$. $F$ and $d$ are two free parameters determining the amplitude and period of
the fringing signal.
We fit this function to the observational residuals using least-squares optimization to obtain the 
optimal fringing parameters in each spectral order and fiber.
Then, we applied this optimized fringing formula to the full spectral model during retrieval analysis. 
We note that the fringing signals do not bias the measurement of chemical composition as we obtained consistent results either with or without the treatment of fringing.

\subsection{Retrieval Framework}

Given the model setup detailed in Section~\ref{sec:forward}, we compute synthetic spectra 
of the companion using the radiative transfer code \texttt{petitRADTRANS} \citep[pRT,][]{MolliereEtAl2019}. 
The model accounts for the Rayleigh scattering of H$_2$ and He, the collision-induced absorption of H$_2$-H$_2$ and H$_2$-He, and
the scattering and absorption cross sections of crystalline, irregularly shaped \enstatite~cloud particles. 
We use the line-by-line mode of \texttt{pRT} to calculate the emission spectra at high spectral resolution. 
To speed up the calculation, we downsample the original opacity tables (with $\lambda/\Delta\lambda \sim10^6$) by a factor of 3. This downsampling factor has been tested to ensure that it does not bias the retrieval results \citep{ZhangEtAl2021a, XuanEtAl2022}.
We include opacity from \HTWOO~\citep{PolyanskyEtAl2018}, \COmain, \COiso~\citep{LiEtAl2015}, \methane~\citep{HargreavesEtAl2020}, 
\COTWO~\citep{RothmanEtAl2010}, and \ammonia~\citep{ColesEtAl2019} 
in our model.

Subsequently, the synthetic high-resolution spectrum is radial-velocity (RV) shifted 
by the systemic and barycentric velocity, and rotationally broadened by $v\sin i$ using the method from
\cite{CarvalhoJohns-Krull2023}.
The model spectrum at native resolution is then convolved with a Gaussian kernel and 
binned to the wavelength grid of the observed spectrum
to match the resolving power of the instrument ($\lambda/\Delta\lambda \sim35,000$). 
The width of the LSF is measured from the point spread function of the stellar observations
in the cross-dispersion direction.
We apply a scaling factor $\mathcal{R}_\mathrm{scaling}$ to the width to account for 
potential inaccuracy in the measured values.
Then, the model spectrum is multiplied by the telluric transmission and instrument response, 
which is obtained by dividing the spectrum of the featureless A-type primary star by a 
PHOENIX stellar model \citep{HusserEtAl2013}.
As the spectrum is contaminated by low-order stellar speckle noise, we discard the continuum information by 
high-pass filtering of both the observation and model using a 150-pixel ($\sim$3 nm) wide Gaussian filter. We note that the retrieval results are robust to the choice of the kernel width.

To construct the full forward model and compare it directly to observations, we took the linear combination of the companion's component $\mathbf{m}_p$ with 
the observed stellar spectrum $\mathbf{m}_\ast$ following \cite{RuffioEtAl2019, WangEtAl2021, LandmanEtAl2024}. 
The full model is calculated as follows:
\begin{equation}
    \mathbf{M} = \mathbf{m}\, \mathbf{c} =
    \begin{bmatrix}
        \mathbf{m}_\ast & \mathbf{m}_p
    \end{bmatrix}
    \begin{bmatrix}
        \mathbf{c}_\ast \\
        \mathbf{c}_p
    \end{bmatrix},
\end{equation}
where the coefficients $\mathbf{c}$ of the two linear components 
can be analytically marginalized by solving the linear equation
\begin{equation}
    \mathbf{d} = \mathbf{m}\, \mathbf{c} + \eta,
\end{equation}
with $\mathbf{d}$ being the observed spectrum and $\eta$ the random noise.
The least square solution for the linear equation can be found by solving
\begin{equation}
    \mathbf{c}^T \mathbf{m}^T \Sigma_0^{-1} \mathbf{m} = \mathbf{d}^T \Sigma_0^{-1} \mathbf{m},
\end{equation}
where $\Sigma_0$ is the covariance matrix with the diagonal items populated by observation uncertainties.
Here, we do not account for the non-diagonal covariance matrix because of the low S/N of the observations. We test for the effects of correlated errors by modeling them with Gaussian Processes (GP) following \cite{deRegtEtAl2024}, and find no difference in the constraints on the free parameters (see Fig.~\ref{fig:corner_gp} in Appendix). This demonstrates that the uncorrelated noise dominates the uncertainty.

The log-likelihood of the model $\mathbf{M}$ is formulated as:
\begin{equation}
    \begin{aligned}
        \ln \mathcal{L} = & -\frac{1}{2} \bigg[ N\ln(2\pi)+\ln(|\Sigma_0|)+N\ln(s^2) \\
        & + \frac{1}{s^2}(\mathbf{d} - \mathbf{M})^T \Sigma_0^{-1} (\mathbf{d} - \mathbf{M}) \bigg],
    \end{aligned}
\end{equation}
where $N$ is the number of data points, and $s$ is an error bar inflation parameter, 
which accounts for the underestimated uncertainties of the observations. Following \cite{RuffioEtAl2019}, 
the optimal $s$ for each model can be calculated by 
effectively scaling the reduced $\chi^2$ to 1:
\begin{equation}
    s^2 = \frac{1}{N} (\mathbf{d} - \mathbf{M})^T \Sigma_0^{-1} (\mathbf{d} - \mathbf{M}).
\end{equation}
The error bar inflation parameter can essentially account for the model's systematic errors compared to the observations, which ensures that these uncertainties are propagated to the posteriors of the free parameters.
The log-likelihood is evaluated in each spectral order and fiber before being combined, 
allowing for different linear coefficients and error bar inflation factors optimized for each order.

In total, the retrieval with disequilibrium chemistry model has 22 free parameters and the free chemistry model has 23 free parameters, as summarized in Table~\ref{tab:params}.
For the Bayesian inference process, we use the nested sampling tool \texttt{PyMultiNest} \citep{BuchnerEtAl2014}, 
which is a Python wrapper of the \texttt{MultiNest} method \citep{FerozEtAl2009}.
The retrievals are performed in importance nested sampling mode with a constant efficiency of 5\%. 
It uses 2000 live points to sample the parameter space and derives the posterior distribution of free
parameters.


\begin{deluxetable}{cccc}
\tabletypesize{\scriptsize}
    \tablecaption{Priors and posteriors of the retrievals.}
    \tablehead{\colhead{Parameter} & \colhead{Prior} & \colhead{Disequilibrium} & \colhead{Free}}
    \startdata
    $M_p$ (\Mjup) & $\mathcal{N}$(16.1, 5.0) & $17.2 \pm 4.2$ & $17.3 \pm 4.2$\\
    $R_p$ ($R_\mathrm{Jup}$) & $\mathcal{U}$(0.7,\ 1.3) & $0.94_{-0.16}^{+0.21}$ & $0.93_{-0.15}^{+0.22}$\\
    $v\sin i$ \kms & $\mathcal{U}$(0.1,\ 20) & $3.5_{-1.9}^{+1.7}$ & $3.4_{-1.9}^{+1.7}$\\
    RV \kms & $\mathcal{U}$(-30,\ 30) & $17.9 \pm 0.3$ & $17.9 \pm 0.3$ \\
    $\epsilon_\mathrm{limb}$ & $\mathcal{U}$(0,\ 1)& $0.5 \pm 0.3$ & $0.5 \pm 0.3$ \\
    $\mathcal{R}_\mathrm{scaling}$ & $\mathcal{U}$(0.8,\ 1.2)& $0.96_{-0.10}^{+0.12}$  & $0.95_{-0.10}^{+0.12}$ \\
    $\rm [M/H]$ & $\mathcal{U}$(-1.5,\ 1.5) & $0.26_{-0.23}^{+0.24}$ & $0.22 \pm 0.23$  \\
    C/O & $\mathcal{U}$(0.1,\ 1.5) & $0.55_{-0.04}^{+0.06}$ & $0.64 \pm 0.05$\\
    log$X^\mathrm{H_2O}$ & $\mathcal{U}$(-12,\ -1) & - & $-2.47 \pm 0.21$ \\
    log$X^\mathrm{CO}$ & $\mathcal{U}$(-12,\ -1) & - & $-2.03 \pm 0.23$ \\
    log$X^\mathrm{CH_4}$ & $\mathcal{U}$(-12,\ -1) & - & $-5.2_{-4.4}^{+0.6}$ \\
    log$X^\mathrm{NH_3}$ & $\mathcal{U}$(-12,\ -1) & - &  $-7.5_{-2.9}^{+2.5}$\\
    log$X^\mathrm{^{13}CO}$ & $\mathcal{U}$(-12,\ -1) & - & $-4.5_{-4.2}^{+0.6}$ \\
    log(\COiso/\COmain) & $\mathcal{U}$(-12,\ -1) & $-2.7_{-5.6}^{+0.8}$ & - \\
    log$P_\mathrm{quench}$ (bar) & $\mathcal{U}$(-5,\ 2)  & $0.21_{-0.83}^{+1.06}$ & - \\
    $T_0$ (K) & $\mathcal{U}$(1500,\ 5000) & $3904 \pm 785$ & $4094 \pm 724$ \\
    $(d\ln T /d\ln P)_0$  & $\mathcal{U}$(0.02,\ 0.05) & $0.035 \pm 0.01$ & $0.035 \pm 0.01$ \\
    $(d\ln T /d\ln P)_1$  & $\mathcal{U}$(0.03,\ 0.07) & $0.05 \pm 0.01$ & $0.05 \pm 0.01$ \\
    $(d\ln T /d\ln P)_2$  & $\mathcal{U}$(0,\ 0.15) & $0.07 \pm 0.05$ & $0.08 \pm 0.05$ \\
    $(d\ln T /d\ln P)_3$  & $\mathcal{U}$(0,\ 0.5) & $0.22 \pm 0.09$ &  $0.29 \pm 0.12$ \\
    $(d\ln T /d\ln P)_4$  & $\mathcal{U}$(0.,\ 0.5) & $0.12 \pm 0.05$ & $0.13 \pm 0.06$ \\
    $(d\ln T /d\ln P)_5$  & $\mathcal{U}$(0.,\ 0.5) & $0.22 \pm 0.10$ & $0.24 \pm 0.09$ \\
    $(d\ln T /d\ln P)_6$  & $\mathcal{U}$(0,\ 0.5) & $0.20 \pm 0.15$ & $0.17 \pm 0.14$ \\
    log($X_0^\mathrm{MgSiO_3}$)& $\mathcal{U}$(-12,\ -1) & $-7.4\pm 3.0$ & $-7.4\pm 3.0$ \\
    $f_\mathrm{sed}$ & $\mathcal{U}$(0,\ 10) & $6.0 \pm 2.5$ & $6.0 \pm 2.5$\\ 
    log($K_\mathrm{zz}$) & $\mathcal{U}$(5,\ 13) & $6.0 \pm 2.6$ & $6.0 \pm 2.6$ \\
    $\sigma_g$ & $\mathcal{U}$(1.05,\ 3) & $2.0 \pm 0.6$ & $2.0 \pm 0.6$\\
    \enddata
    
    \label{tab:params}
    \tablecomments{$\mathcal{U}$(a, b) represents a uniform distribution and 
    $\mathcal{N}$(a, b) represents a normal distribution. The last two columns show
    posteriors with 1$\sigma$ uncertainties from the disequilibrium chemistry model and
    free chemistry model.}
\end{deluxetable}

\section{Retrieval results} \label{sec:result}

We jointly fit two epochs of KPIC observations by generating the model spectrum and calculating the likelihood for each epoch and summing up the two likelihood values.  
We show retrieval results for two model setups: disequilibrium and free chemistry.
The posterior distributions of the free parameters in these models are summarized in Table~\ref{tab:params}.
In general, these models result in consistent constraints on the atmospheric properties.
We also carry out analyses for the two epochs independently. The retrieved parameters are consistent within 1-2$\sigma$, as shown in Fig~\ref{fig:compare_epochs}. The first epoch provides slightly tighter constraints because of its better S/N (see Table~\ref{tab:obs}).

\subsection{Cross-correlation detection}

\begin{figure}[t!]
    \includegraphics[width=\linewidth]{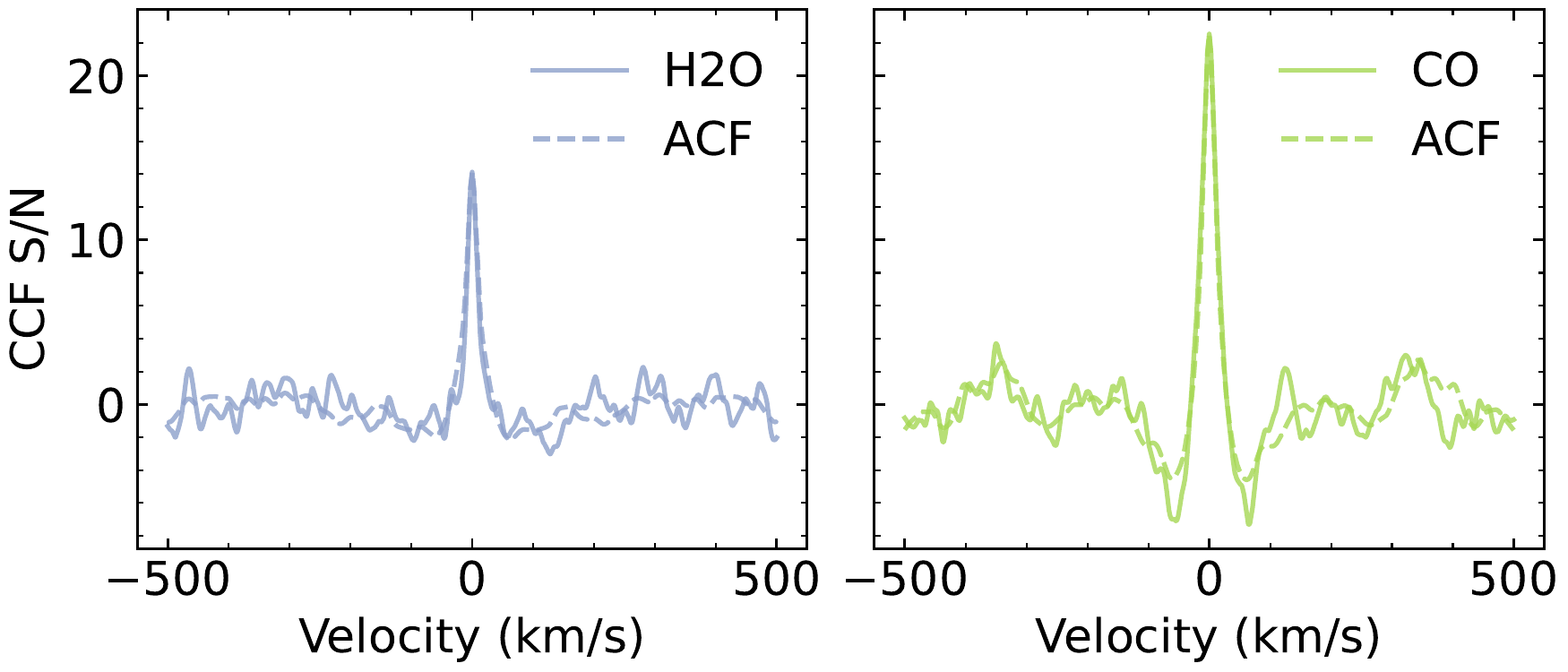}
    \caption{Cross-correlation detection of \HTWOO\ and CO in KPIC observations of \target. 
    The cross-correlation functions (CCF) are scaled by the noise to show the detection S/N on the y-axis.  
    The dashed lines show the auto-correlation function (ACF) of each molecular template.
    \label{fig:ccf}}
\end{figure}

We carried out a cross-correlation analysis to show the detection of molecules in \target.
The cross-correlation functions (CCF) were computed as 
\begin{equation}
        \mathrm{CCF}(v) = \frac{1}{s^2} \mathbf{F}(v) ^T  \Sigma_0^{-1} \mathbf{R},
\end{equation}
where $\mathbf{R}$ is the observed spectra minus the best-fit disequilibrium chemistry model with the abundance of 
a specific molecule being set to zero; $\mathbf{F}$ is the molecular template which was computed 
by differencing the best-fit model and the best-fit model without the contribution of that molecule;
$v$ is the radial velocity shift between the model and data.
Both $\mathbf{R}$ and $\mathbf{F}$ were high-pass filtered using a median filter with a width of 100 pixels.
Then, the CCFs of individual orders were combined into a master CCF, as shown in Fig.~\ref{fig:ccf}.
The noise of the CCF was estimated by subtracting the model's auto-correlation function (ACF) and taking the standard deviation at $|v|>150$ \kms.
We detected \HTWOO\ and CO with an S/N of 14 and 23, respectively. 
We checked other molecules, including \methane, \ammonia, \COiso, H$_2$S, and CO$_2$, and found no significant detections (see Fig.\ref{fig:non-detection} in Appendix).

\begin{figure*}[t!]
    \includegraphics[width=\linewidth]{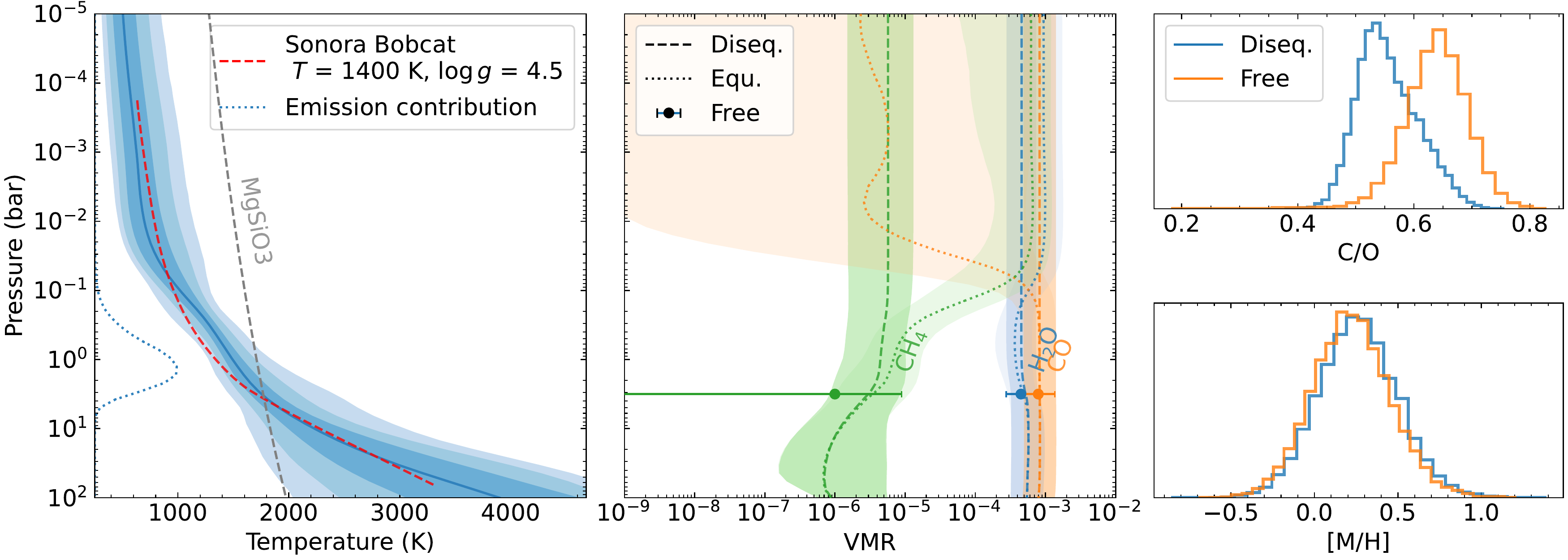}
    \caption{Retrieved temperature structure and chemical abundances of \target. 
    Left panel: best-fit T-P profile with 1$\sigma$, 2$\sigma$, and 3$\sigma$ envelope are shown in blue.
    The red dashed line is self-consistent T-P profile of $T\mathrm{eff}=1400$ K and $\log g =4.5$
    from the Sonora model \citep{MarleyEtAl2021}. The blue dotted line represents the flux-weighted 
    emission contribution of the model. The gray dashed line shows the condensation curve 
    of the \enstatite~cloud. The middle panel shows retrieved volume mixing ratios (VMRs) of the major 
    molecules with 1$\sigma$ envelopes. The dashed lines show the results of the disequilibrium chemistry model;
    the dotted lines represent the equilibrium case when the chemical quenching is manually turned off.
    The error bars denote the retrieved VMRs in the free chemistry model.
    The right panels compare the retrieved C/O and metallicity of the disequilibrium and free chemistry models. 
    The free chemistry model retrieves a slightly higher C/O ratio is because of the hidden oxygen in cloud condensates, such as enstatite.
    \label{fig:chem}}
\end{figure*}

\subsection{C/O and Metallicity}

The retrieved chemical abundances and metallicity [M/H] and C/O ratios using two models are 
shown in Fig.~\ref{fig:chem}.
Using the disequilibrium chemistry model, we obtained a C/O=$0.55_{-0.04}^{+0.06}$ and 
[M/H]=$0.26_{-0.23}^{+0.24}$ (1$\sigma$ confidence intervals).
The empirical level of systematic uncertainty of KPIC measurements is $\sim$ 20\% for C/O ratio and 0.2 dex for metallicity,
as estimated from benchmark brown dwarf companions, HR 7672 B and HD 4747 B \citep{WangEtAl2022, XuanEtAl2022}. These are roughly consistent with our results on \target.
The free chemistry model resulted in a derived C/O of $0.64 \pm 0.05$ and metallicity of $0.22 \pm 0.23$, 
in line with the disequilibrium chemistry model.
In comparison, the free chemistry model retrieved 
consistent VMRs for the major molecules as shown in Fig.~\ref{fig:chem}. 
The slightly different C/O ratio derived from the free chemistry 
model can be attributed to the cold trapping of oxygen in clouds \citep{WoitkeEtAl2018}. 
Condensates such as \enstatite\ take a fraction of oxygen out of the gas phase.
The chemical model predicts a mass fraction of $\sim 2.4\times 10^{-3}$ for \enstatite, which was not constrained with our analysis (see Section~\ref{sec:cloud}).
Taking this amount of clouds into account, we estimated that the actual C/O ratio in the free chemistry model would be $\sim 0.55$, consistent with the result from the disequilibrium chemistry model. 

Regarding Bayesian evidence, the disequilibrium chemistry model slightly outperforms the free chemistry model
at 3$\sigma$ significance \citep{BennekeSeager2013}. The values are summarized in Table~\ref{tab:evidence}.
We found that the log likelihood of the best-fit disequilibrium model is the same as the free chemistry model. 
This suggests that both models fit the observations equally well, and the slightly lower Bayesian evidence of the free chemistry model is due to the increased number of free parameters. 
Additionally, we tested imposing a Gaussian prior of $1.1\pm0.1$ \Rjup~on the radius of the companion following the evolutionary model \citep{BaraffeEtAl2003, CurrieEtAl2023} 
and found that the retrieval results remained entirely consistent.


\subsection{Disequilibrium Chemistry}

The disequilibrium chemistry model retrieves a quench pressure of 
$P_\mathrm{quench} \sim 0.2 - 22.4$ bar (1$\sigma$).
This quench pressure is needed in order to explain the observed under-abundance 
of \methane\ relative to CO in the atmosphere. 
In contrast, turning off chemical quenching results in a transition from CO-dominated
to a \methane-dominated composition at pressures below 0.1 bar given the T-P profile 
(see Fig.\ref{fig:chem}). 
The KPIC observations do not support this equilibrium case, as the VMR of CO in our free chemistry retrieval is constrained to a higher 
value than the equilibrium prediction and we obtain a 3$\sigma$ upper limit of $<4\times10^{-5}$ on the VMR of \methane.

We re-ran the retrieval assuming chemical equilibrium without quenching and compared the resulting fit to the nominal model with quenching. 
The equilibrium retrieval converged to a more isothermal T-P profile (i.e., hotter upper layers)
in order to avoid the transition to a \methane-dominated atmosphere.
We calculated the Bayes factor for these two models and found that the nominal
disequilibrium chemistry model
is favored over that for equilibrium at a 2.4$\sigma$ significance. 
Similar findings have been observed in several late L-type super-Jupiters 
such as HR 8799 cde \citep[e.g.][]{KonopackyEtAl2013, BarmanEtAl2015, MolliereEtAl2020} and brown dwarfs 
\citep[e.g.][]{XuanEtAl2022, deRegtEtAl2024}. 


\subsection{Clouds} \label{sec:cloud}

The cloud-related parameters were not constrained in our retrieval analysis 
(see Table~\ref{tab:params} and Fig.~\ref{fig:corner}). 
We calculated the Bayesian evidence for the clear versus cloudy models and found the clear model is slightly preferred at 2.7$\sigma$ significance because of fewer free parameters.
We retrieved consistent chemical abundances from the cloudy and cloud-free models.
While clouds are expected in atmospheres of late L-type brown dwarfs \citep{BurrowsEtAl2006}, 
the reason for our data's insensitivity to clouds may be twofold.
First, the limited S/N of the observations may not be sufficient to distinguish the small differences in line contrasts
caused by the cloud opacity. 
Second, the silicate cloud base in our model is located at $\sim$10 bar, below the photosphere of 
the atmosphere in K band, as shown in Fig.~\ref{fig:chem}. 
Hence, the impact of cloud opacity is not significant at the 
pressure levels probed by the high-resolution K-band observations 
\citep[see also][]{ZhangEtAl2021a, XuanEtAl2022, LandmanEtAl2024, InglisEtAl2024}.


\begin{figure}[t!]
    \includegraphics[width=\linewidth]{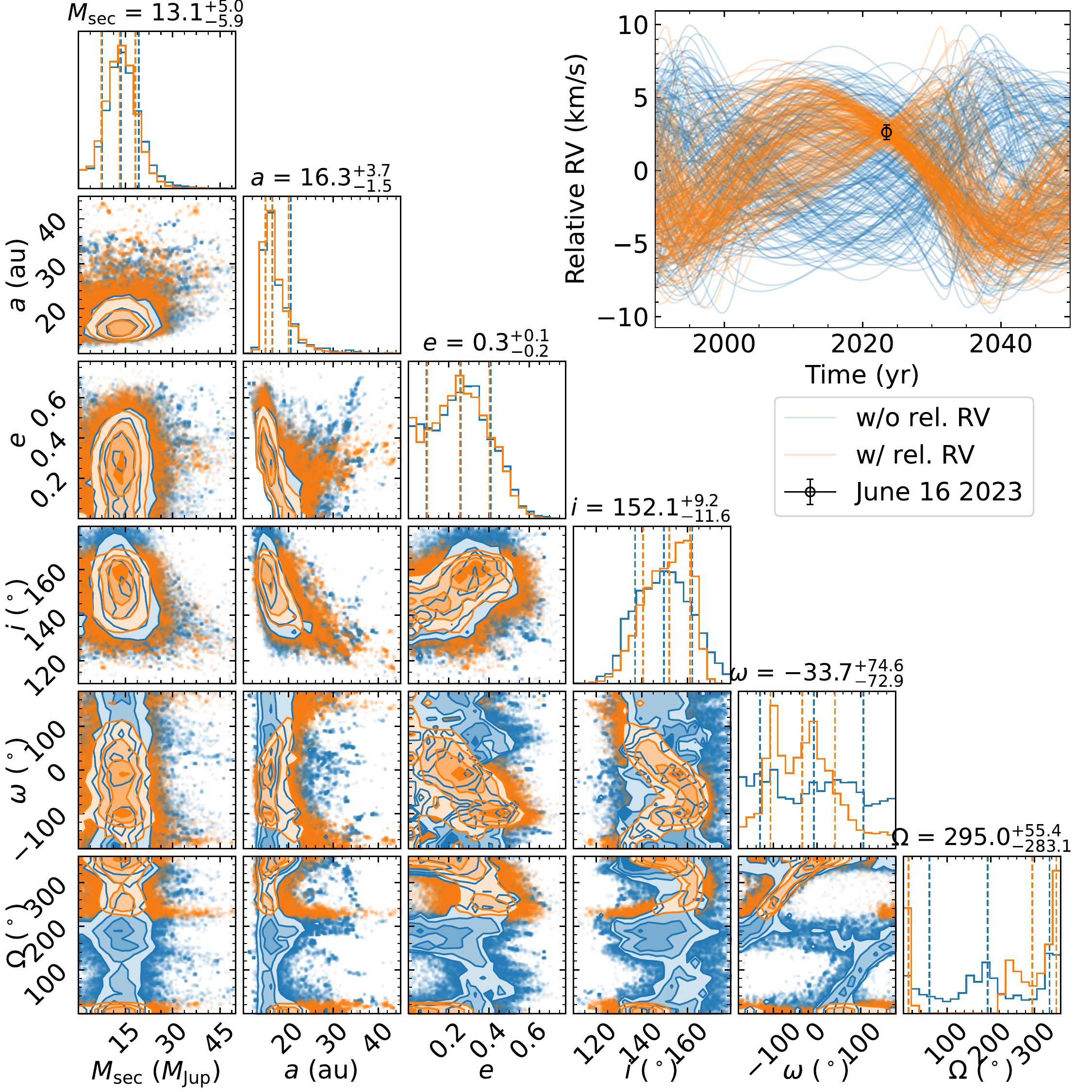}
    \caption{Comparison of the orbital fitting of the \primary~system with and without the relative RV measurement 
    in orange and blue, respectively.
    The corner plot shows posterior distributions of parameters including 
    companion's mass $M_\mathrm{sec}$, semimajor axis $a$, eccentricity $e$, inclination $i$, argument of periastron $\omega$, and longitude of ascending node $\Omega$.
    The upper right panel shows the relative RV over time from random draws of the orbital solutions 
    in both fits. Adding our RV measurement to the fitting helps improve constraints on $i$, $\omega$, and $\Omega$. The updated 1$\sigma$ constraints are labeled on top of each column. 
    \label{fig:relRV}}
\end{figure}

\subsection{Rotation and RV}

The projected rotation velocity of \target~is measured to be $v\sin(i) = 3.5_{-1.9}^{+1.7}$ \kms\ in our retrievals. 
However, considering the spectral resolution of KPIC ($\mathcal{R}\sim 35,000$), 
the instrument broadening dominates over the rotational broadening of the target, 
meaning that the rotation is not resolved in our data. 
The likelihood of the best-fit retrieval model shows no difference whether including spin as a free parameter or not.
Therefore, we report a 3$\sigma$ upper limit of $v\sin(i)<7.8$ \kms~as derived from the posterior distribution. This is also consistent with the spectral resolution of KPIC. 
Adopting an isotropic distribution of inclination, we estimated its spin velocity to be $<10$ \kms, which corresponds to a rotation velocity versus break-up velocity $v/v_\mathrm{break}\sim0.06$.
We further discuss this constraint in Section~\ref{sec:spin}.

We measured a radial velocity of $-17.9\pm 0.3$ \kms for HIP 99770 b. This allows us to compute the relative RV 
between the star and companion and add to the orbital constraints of the system. 
Since the A-type host star is almost featureless in K band, we cannot measure the stellar RV in our KPIC data.
The stellar RV is not fully consistent in the literature, likely because 
the fast rotation of the star \citep[$\sim 80$ \kms,][]{GaiaCollaborationEtAl2021} makes it challenging to measure the RV accurately. 
We adopted the RV measurement of \primary~from Gaia DR3 of $-20.5\pm0.4$ \kms~\citep{GaiaCollaborationEtAl2021}. This led to a relative 
RV of $2.6\pm0.5$ \kms~between the companion and primary.
Adding this relative RV constraint, we carried 
out orbital fitting using the \texttt{orvara} code \citep{BrandtEtAl2021a} following \citep{CurrieEtAl2023}
to refine the orbital properties of \target. 
We note that here we used the default 1/$M_\mathrm{sec}$ prior for the secondary mass, 
which is responsible for the lower $M_\mathrm{sec}$ ($\sim13$ \Mjup) as was also mentioned in \cite{CurrieEtAl2023}.
We compared the posteriors of the orbital fitting with and without the relative RV measurement in Fig.~\ref{fig:relRV}.
Although the addition of relative RV did not significantly change the overall orbital constraints, 
it assisted in ruling out some solutions and improving the constraint on the argument of periastron $\omega$, longitude of ascending node $\Omega$, and orbital inclination $i= 152_{-11}^{+9}$ deg. 
Additional epochs of RV measurements in five-year baseline will 
further strengthen the constraining power
on other parameters, such as the secondary mass and orbital eccentricity.

\section{Discussion} \label{sec:discussion}

\subsection{Rotation velocity} \label{sec:spin}

\begin{figure}[t!]
    \includegraphics[width=\linewidth]{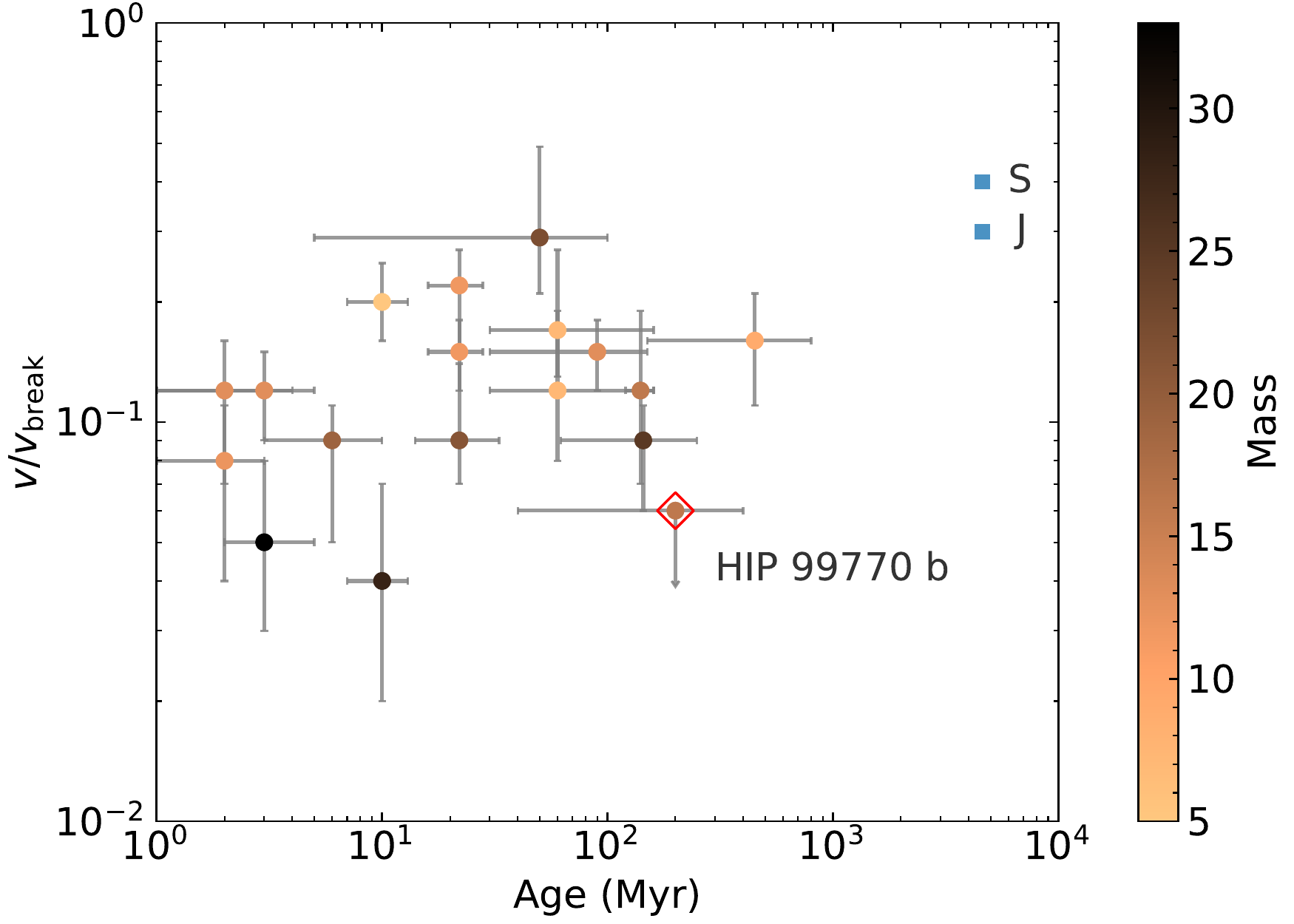}
    \caption{Rotation velocities (assuming isotropic inclination) as a fraction of break-up velocities for super-Jovian companions.
    Our measurement of \target\ is denoted with a red diamond.
    The measurements of other companions are adopted from the literature, including 
    \cite{BryanEtAl2020, WangEtAl2021,XuanEtAl2024b}. The data are color-coded by the companion masses.
    The blue data points show the spin of Jupiter and Saturn in the solar system.
    \label{fig:spin}}
\end{figure}

Putting our upper limit on rotation velocity into context, we note that the spin of \target~is at the low end
compared to literature measurements of other super-Jovian companions as shown in Fig.~\ref{fig:spin}.
Previous studies suggest that younger companions generally have lower rotation rates, 
which are expected to 
increase with age as their radii contract following angular momentum conservation \citep{BryanEtAl2020, VosEtAl2020}. 
In contrast, field brown dwarfs display a larger scatter in spin \citep{HsuEtAl2021}.
Although the age of \target~is older than 40 Myr, its rotation velocity (assuming isotropic inclination distribution) 
is comparable to that of very young companions ($<10$ Myr) such as GQ Lup b \citep{SchwarzEtAl2016}, 
DH Tau b, and HIP 79098 b \citep{XuanEtAl2024b}, and is lower than those at older ages.
If the trend of increasing spin with increasing age holds, it may indicate a nearly pole-on orientation for \target. 
If we assume that its spin and orbit are aligned ($i\sim152$), the spin velocity is $<15$ \kms~or $v/v_\mathrm{break} \sim 0.09$, 
making it more comparable to other super-Jovian companions. 

On the other hand, the slow spin may be indicative of effective magnetic braking by the circumplanetary disk (CPD)
\citep{Batygin2018, GinzburgChiang2020, WangEtAl2021}. 
Massive companions are expected to effectively ionize the CPD 
and interactions with magnetic fields then act to spin down the companions.
Therefore, the slow spin of \target~may 
indicate that it hosted an unusually long-lived CPD and/or that it formed early via gravitational instability, 
both of which allow for a longer time for 
the companion to spin down.
The large scatter in the measured spins of super-Jupiter companions may represent the consequences of 
mixed formation pathways and histories \citep{BryanEtAl2020}.
Future rotation period measurements with light curves will help break this degeneracy with inclination.
More accurate measurements of rotation velocities using high-resolution, large spectral grasp instruments such as
VLT/CRIRES$^+$ and Keck/HISPEC will be essential for understanding the population-level trend.

\subsection{Implication for Formation}

We retrieve an atmospheric [M/H] $=0.26 \pm 0.24$ and C/O $=0.55 \pm 0.06$ (1$\sigma$ confidence intervals) for \target, consistent with the solar values. 
Previous studies reported stellar abundances of [C/H]$\sim$0.18 $\pm$ 0.09 and [O/H]$\sim$0.01 $\pm$ 0.09 
\citep{ErspamerNorth2003, HinkelEtAl2014}. 
This corresponds to a C/O ratio of $0.8_{-0.2}^{+0.3}$, roughly consistent with solar to super-solar value of $0.59 \pm 0.08$ \citep{AsplundEtAl2021}.
The uncertainties in the stellar composition result from the large rotational broadening of the primary and the deficit of oxygen lines in its spectrum. 
A more accurate stellar C/O constraint is needed in order to draw unambiguous conclusion on the formation of the companion.
For consistency, we simply assume a solar composition for the primary star \primary~in the following discussion.

The companion's current semimajor axis of 17 au is well within the typical location of the CO iceline 
in protoplanetary disks around early-type stars \citep{QiEtAl2015}.
This makes it an interesting case for testing formation models.
Given our constraints on its atmospheric composition,
the companion is compatible with a formation via either core accretion or gravitational instability.
Based on the static picture of chemistry in smoothed protoplanetary disks \citep{ObergEtAl2011},
the core accretion scenario is expected to result in either: 1) a substellar C/O and metal-rich atmosphere 
because of the accretion of
oxygen-enhanced solids,  2) an elevated C/O combined with substellar metallicity 
due to gas-dominated accretion, or 3) a stellar C/O when the accreted material corresponds to an overall stellar metallicity for the atmosphere. 
In practice, protoplanetary disks are not static and have complicated substructures \citep{Oberg2021}. Dust drift and chemical evolution of gas and dust can alter the distribution of disk elemental abundances over time \citep{MolliereEtAl2022a}. Planet formation models coupled with disk chemistry evolution and dust migration are needed to investigate these questions further. Qualitatively, disk observations support the simplified picture that the gas-phase oxygen is strongly depleted as more O-bearing species are locked in the ice phase beyond the snowline \citep{LeGal2021, Bosman2021}. Therefore, planets formed under such conditions are expected to follow the scenarios outlined above.

In contrast, gravitational instability should lead to similar stellar and companion atmospheres
since it is thought to occur relatively early when the disk is massive and the reservoir of solids is pristine \citep{SchibEtAl2021}. 
Although the companion may undergo late enrichment by solid accretion after the initial collapse, 
this would not significantly alter the composition due to the high envelope mass; for example,
to enhance the metallicity by one dex would require accreting $\sim$2 \Mjup~of solids 
\citep{InglisEtAl2024}.
Similarly, we do not expect a large enrichment of metal in super-Jupiter atmospheres in core accretion models.
Therefore, the solar composition of \target~does not preclude either formation scenarios. 
If the companion was born ex-situ, it may have formed in the outer disk
followed by type II migration \citep{DurmannKley2015, RobertEtAl2018}. 
Such planet migration could potentially boost the accretion of planetesimals, leading to a slight metal enrichment, as we constrain for \target.


The chemical composition of \target~also fits in the overall trend for super-Jupiter 
companions. 
\cite{XuanEtAl2024b} carried out retrieval analyses in a sample of 
10-30 \Mjup~companions using KPIC observations. They found the companions generally had
solar compositions, hinting at formation via gravitational collapse or plausible core 
accretion beyond CO iceline. In contrast, lower mass ($<10$ \Mjup) companions appear to display
a larger range in C/O and metallicity \citep[e.g,][]{ZhangEtAl2023, WhitefordEtAl2023, LandmanEtAl2024, NasedkinEtAl2024}. 
By comparing the C/O ratio in a sample of directly imaged companions with transiting exoplanets, \cite{HochEtAl2023} suggested that there are two distinct populations with a boundary at $\sim$5 \Mjup, implying a transition of
formation pathways near a mass range of $5-10$\Mjup. 
However, more work is needed to determine whether there exists a clear mass boundary for distinct formation channels, as such compositional measurements get challenging towards lower-mass and closer-in companions.
The comparison between directly imaged super-Jupiters and transiting exoplanets with similar masses will be informative in distinguishing whether these populations are separated by different formation pathways or migration histories \citep{KemptonKnutson2024}.
It also calls for the joint understanding of trends in the orbital architectures of these systems \citep{NielsenEtAl2019, BowlerEtAl2020, NagpalEtAl2023, DoOEtAl2023}
to unravel their formation.

\section{Conclusion}\label{sec:conclusion}

We carried out detailed spectral characterization of the recently discovered super-Jupiter 
companion \target~using the fiber-fed high resolution spectrograph KPIC ($\mathcal{R}\sim 35,000$). 
The K-band observations led to clear detections of \HTWOO\ and CO in the atmosphere of \target.
We used atmospheric retrievals to constrain its atmospheric composition, including its C/O and metallicity, 
projected rotation velocity ($v\sin i$), and radial velocity (RV). 

\begin{itemize}
    \item We found the companion's atmosphere has C/O $=0.55 \pm 0.06$ and [M/H] $=0.26\pm 0.24$ (1$\sigma$ confidence interval), which are 
    consistent with stellar values. 
    It joins the ensemble of super-Jovian companions showing broadly solar compositions.
    This is compatible with a formation 
    via either gravitational instability or core accretion. 
    \item We found that the ratio of \methane~to CO in the companion's atmosphere is lower than predicted by equilibrium chemistry models, and place a lower bound of 0.2 bars on the quench pressure.
    \item Although the companion is expected to be cloudy, we found that our models do not require clouds. This may be because of the limited S/N, or or because the clouds are located below the photosphere for our $K$ band observations.  
    The presence or absence of clouds could be better constrained by expanding the wavelength range of these observations, which would increase our sensitivity to the wavelength-dependent scattering signal from clouds.
    \item We added the companion-to-primary relative RV measurement to the orbital fitting and obtained updated
    constraints on the companion's orbital $i$, $\omega$, and $\Omega$.
    \item The projected rotation velocity
     $v\sin(i)<7.8$ \kms\ is small compared to other directly imaged companions with similar ages and masses. 
     This may indicate a nearly pole-on orientation or effective
    magnetic braking by a circumplanetary disk.
\end{itemize}

Located within 20 au and straddling the deuterium-burning mass boundary, \target\ represents an intriguing
target to investigate the link between the present atmosphere and formation history.
The convoluted formation and evolution processes make it challenging to retrace the origin of planets
unambiguously.
Modeling studies on the formation of these well-characterized super-Jovian companions are crucial for advancing 
our knowledge. 
In the future, higher S/N spectroscopic observations will allow for tighter constraints 
on the metallicity 
by probing elements beyond C and O. The estimates of solid budgets will massively benefit the 
inference of formation history \citep{LothringerEtAl2021, ChachanEtAl2023}.
Other formation tracers, such as carbon isotope ratios, will help pin down the formation
location relative to the CO iceline, therefore allowing us to lift the degeneracy 
between formation mechanisms \citep{ZhangEtAl2021}.
Detailed characterization of similar companions will play an essential role in
charting the parameter space to understand the formation of the super-Jovian population.

\begin{acknowledgements}

Y.Z. is thankful for support from the Heising-Simons Foundation 51 Pegasi b Fellowship (grant \#2023-4298).
Y.C. acknowledges support from the Natural Sciences and Engineering Research Council of Canada (NSERC) through the CITA National Fellowship and the Trottier Space Institute through the TSI Fellowship.
Funding for KPIC has been provided by the California Institute of Technology, the Jet Propulsion Laboratory, the Heising-Simons Foundation (grants \#2015-129, \#2017-318, \#2019-1312, \#2023-4598), the Simons Foundation, and the NSF under grant AST-1611623.
The data presented herein were obtained at the W. M. Keck Observatory, which is operated as a scientific partnership among the California Institute of Technology, the University of California and the National Aeronautics and Space Administration. The Observatory was made possible by the generous financial support of the W. M. Keck Foundation. The authors wish to recognize and acknowledge the very significant cultural role and reverence that the summit of Mauna Kea has always had within the indigenous Hawaiian community.  We are most fortunate to have the opportunity to conduct observations from this mountain. 
The computation was carried out on the Caltech High Performance Cluster.

\end{acknowledgements}

\vspace{5mm}
\facilities{Keck (KPIC)}

\software{
\texttt{numpy}~\citep{Harris2020},
\texttt{scipy}~\citep{VirtanenEtAl2020},
\texttt{matplotlib}~\citep{Hunter2007},
\texttt{astropy}~\citep{AstropyCollaborationEtAl2018},
\texttt{petitRADTRANS}~\citep{MolliereEtAl2019}, 
\texttt{PyMultiNest} \citep{BuchnerEtAl2014},
\texttt{orvara}~\citep{BrandtEtAl2021a},
\texttt{corner}~\citep{Foreman-Mackey2016},
}

\appendix

\section{Retrieval results}
\begin{figure}[ht]
    \includegraphics[width=\linewidth]{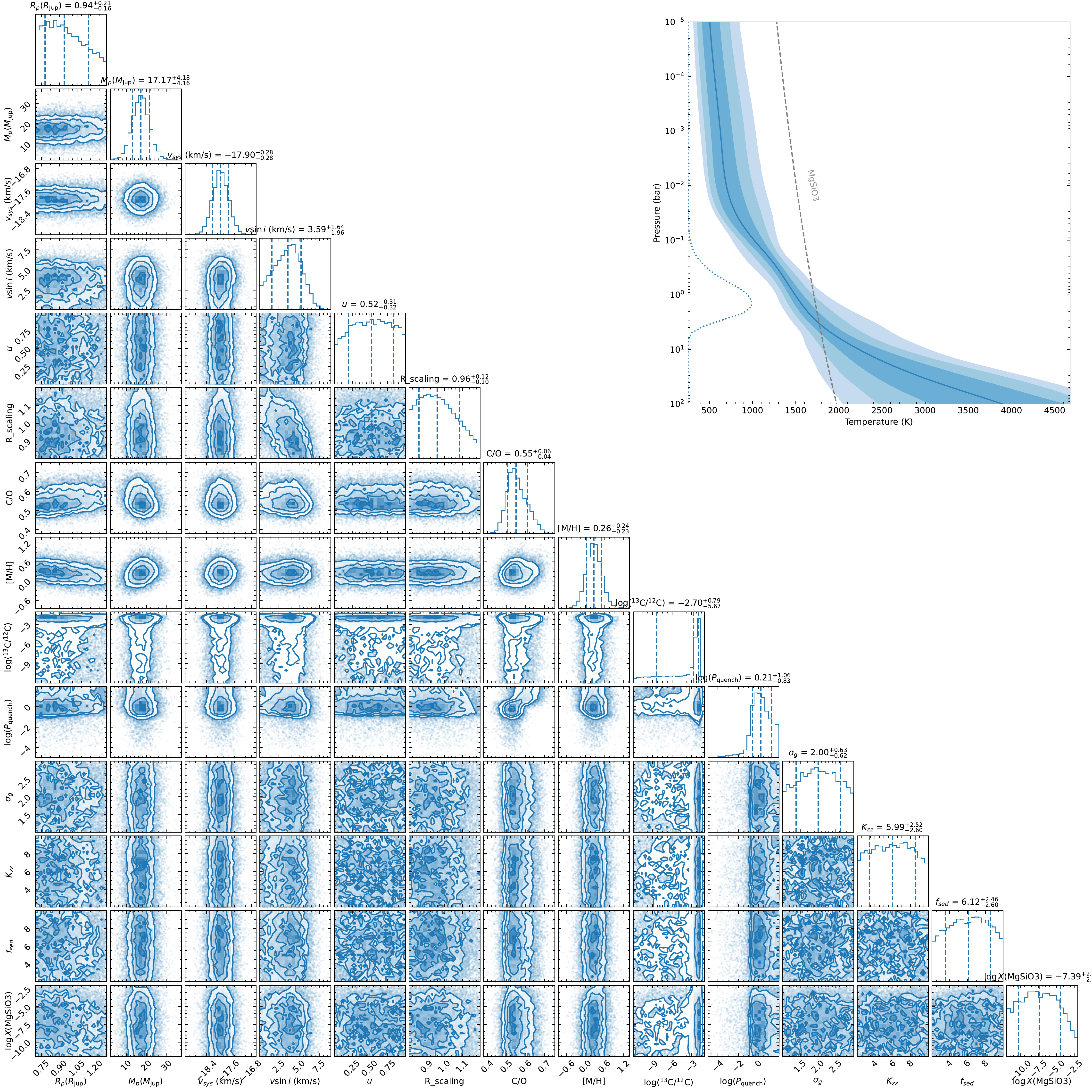}
    \caption{Retrieved posterior distributions of free parameters and temperature-pressure profile 
    in disequilibrium chemistry model using KPIC observations of \target.
    \label{fig:corner}}
\end{figure}

\begin{figure}[ht]
    \includegraphics[width=\linewidth]{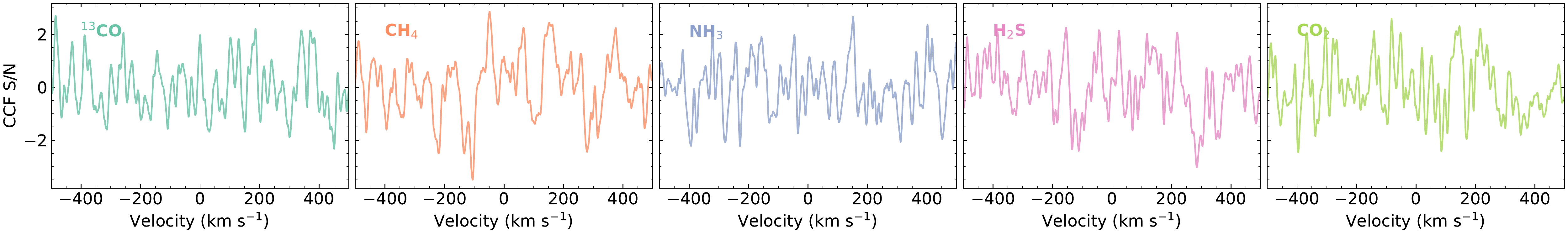}
    \caption{CCFs of non-detected molecules including \COiso, \methane, \ammonia, H$_2$S, and CO$_2$.
    \label{fig:non-detection}}
\end{figure}

\begin{figure}[ht]
    \centering
    \includegraphics[width=0.8\linewidth]{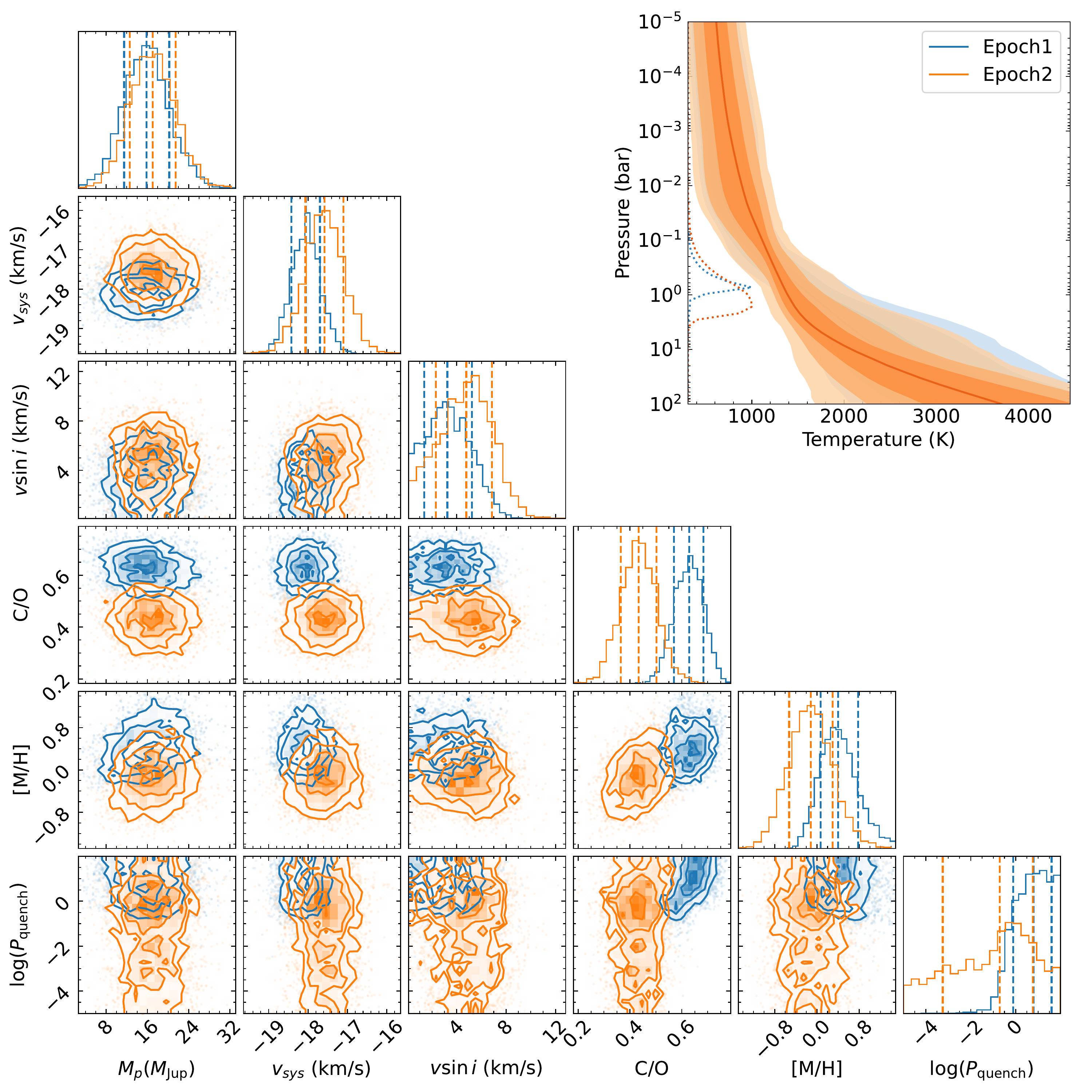}
    \caption{Comparison of independent retrieval analyses on two epochs of observations. The results are consistent within 1-2$\sigma$.
    \label{fig:compare_epochs}}
\end{figure}

\begin{figure}[ht]
    \includegraphics[width=\linewidth]{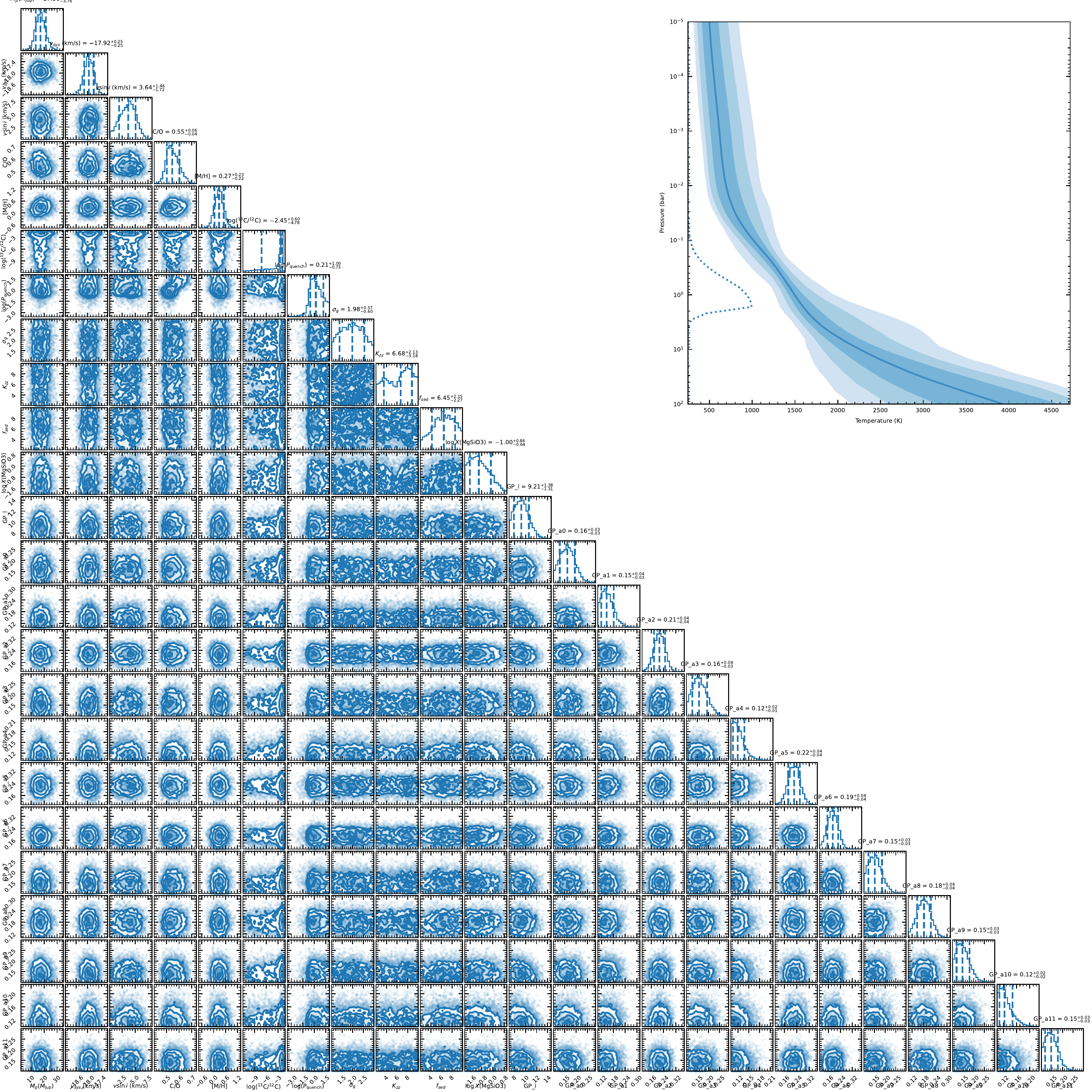}
    \caption{Retrieved posterior distributions of free parameters with Gaussian processes (GP). GP\_l is the length scale of GP in unit of \kms, GP\_a0 to GP\_a11 represents the amplitude of GP (in unit of typical data uncertainties) for each order in each fiber and epoch. The constraints are fully consistent with the model without GP.
    \label{fig:corner_gp}}
\end{figure}

We summarize the Bayesian evidence of various retrieval models mentioned in Section~\ref{sec:result}.
Fig.~\ref{fig:corner} shows the posterior distributions of free parameters in the baseline retrieval with 
disequilibrium chemistry model.
In Table~\ref{tab:evidence}, we list the Bayesian evidence of various retrieval models and their comparisons to the baseline model.
The cross-correlation functions of molecular non-detections are shown in Fig.\ref{fig:non-detection}.
We carry out independent retrievals on two epochs of KIPC data and find broadly consistent posterior constraints as shown in Fig.~\ref{fig:compare_epochs}.
We also explore the effect of correlated noise on the retrieval results by implementing Gaussian Processes. The corner plot of the posteriors is shown in Fig.\ref{fig:corner_gp}. 

\begin{deluxetable}{cccccc}
    \tablecaption{Summary of Bayesian evidence of various retrieval models}
    \tablehead{\colhead{Model} & \colhead{$\ln Z$} & \colhead{$\Delta \ln Z$} & \colhead{Significance ($\sigma$)} 
    }
    \startdata
        Disequilibrium chemistry  & -93748.5 & - & -  \\
        Free chemistry  & -93751.8 & -3.3 & 3.1  \\
        Equilibrium chemistry  & -93750.0 & -1.5 & 2.4  \\
        Disequilibrium chemistry, no cloud  & -93746.2 & 2.3 & 2.7  \\
    \enddata
    \label{tab:evidence}
    \tablecomments{$\Delta \ln Z$ represents the difference of Bayesian evidence $\ln Z$ of each alternative model compared to the nominal disequilibrium chemistry model.}
\end{deluxetable}

\bibliography{HIP99770}

\begin{thebibliography}{}
\expandafter\ifx\csname natexlab\endcsname\relax\def\natexlab#1{#1}\fi
\providecommand{\url}[1]{\href{#1}{#1}}
\providecommand{\dodoi}[1]{doi:~\href{http://doi.org/#1}{\nolinkurl{#1}}}
\providecommand{\doeprint}[1]{\href{http://ascl.net/#1}{\nolinkurl{http://ascl.net/#1}}}
\providecommand{\doarXiv}[1]{\href{https://arxiv.org/abs/#1}{\nolinkurl{https://arxiv.org/abs/#1}}}

\bibitem[{Ackerman \& Marley(2001)}]{AckermanMarley2001}
Ackerman, A.~S., \& Marley, M.~S. 2001, Astrophys. J., 556, 872, \dodoi{10.1086/321540}

\bibitem[{{Agrawal} {et~al.}(2023){Agrawal}, {Ruffio}, {Konopacky}, {Macintosh}, {Mawet}, {Nielsen}, {Hoch}, {Liu}, {Barman}, {Thompson}, {Greenbaum}, {Marois}, \& {Patience}}]{Agrawal2023}
{Agrawal}, S., {Ruffio}, J.-B., {Konopacky}, Q.~M., {et~al.} 2023, \aj, 166, 15, \dodoi{10.3847/1538-3881/acd6a3}

\bibitem[{Asplund {et~al.}(2021)Asplund, Amarsi, \& Grevesse}]{AsplundEtAl2021}
Asplund, M., Amarsi, A.~M., \& Grevesse, N. 2021, Astron. Astrophys., 653, A141, \dodoi{10.1051/0004-6361/202140445}

\bibitem[{{Astropy Collaboration} {et~al.}(2018){Astropy Collaboration}, {Price-Whelan}, Sip{\H o}cz, G{\"u}nther, Lim, Crawford, Conseil, Shupe, Craig, Dencheva, Ginsburg, VanderPlas, Bradley, {P{\'e}rez-Su{\'a}rez}, {de Val-Borro}, Aldcroft, Cruz, Robitaille, Tollerud, Ardelean, Babej, Bach, Bachetti, Bakanov, Bamford, Barentsen, Barmby, Baumbach, Berry, Biscani, Boquien, Bostroem, Bouma, Brammer, Bray, Breytenbach, Buddelmeijer, Burke, Calderone, Cano~Rodr{\'i}guez, Cara, Cardoso, Cheedella, Copin, Corrales, Crichton, D'Avella, Deil, Depagne, Dietrich, Donath, Droettboom, Earl, Erben, Fabbro, Ferreira, Finethy, Fox, Garrison, Gibbons, Goldstein, Gommers, Greco, Greenfield, Groener, Grollier, Hagen, Hirst, Homeier, Horton, Hosseinzadeh, Hu, Hunkeler, Ivezi{\'c}, Jain, Jenness, Kanarek, Kendrew, Kern, Kerzendorf, Khvalko, King, Kirkby, Kulkarni, Kumar, Lee, Lenz, Littlefair, Ma, Macleod, Mastropietro, McCully, Montagnac, Morris, Mueller, Mumford, Muna, Murphy, Nelson, Nguyen, Ninan, N{\"o}the, Ogaz, Oh, Parejko, Parley, Pascual, Patil, Patil, Plunkett, Prochaska, Rastogi, Reddy~Janga, Sabater, Sakurikar, Seifert, Sherbert, {Sherwood-Taylor}, Shih, Sick, Silbiger, Singanamalla, Singer, Sladen, Sooley, Sornarajah, Streicher, Teuben, Thomas, Tremblay, Turner, Terr{\'o}n, {van Kerkwijk}, {de la Vega}, Watkins, Weaver, Whitmore, Woillez, Zabalza, \& {Astropy Contributors}}]{AstropyCollaborationEtAl2018}
{Astropy Collaboration}, {Price-Whelan}, A.~M., Sip{\H o}cz, B.~M., {et~al.} 2018, Astron. J., 156, 123, \dodoi{10.3847/1538-3881/aabc4f}

\bibitem[{Baraffe {et~al.}(2003)Baraffe, Chabrier, Barman, Allard, \& Hauschildt}]{BaraffeEtAl2003}
Baraffe, I., Chabrier, G., Barman, T.~S., Allard, F., \& Hauschildt, P.~H. 2003, Astron. Astrophys., 402, 701, \dodoi{10.1051/0004-6361:20030252}

\bibitem[{Barman {et~al.}(2015)Barman, Konopacky, Macintosh, \& Marois}]{BarmanEtAl2015}
Barman, T.~S., Konopacky, Q.~M., Macintosh, B., \& Marois, C. 2015, Astrophys. J., 804, 61, \dodoi{10.1088/0004-637X/804/1/61}

\bibitem[{Batygin(2018)}]{Batygin2018}
Batygin, K. 2018, Astron. J., 155, 178, \dodoi{10.3847/1538-3881/aab54e}

\bibitem[{Benneke \& Seager(2013)}]{BennekeSeager2013}
Benneke, B., \& Seager, S. 2013, Astrophys. J., 778, 153, \dodoi{10.1088/0004-637X/778/2/153}

\bibitem[{Bitsch {et~al.}(2019)Bitsch, Izidoro, Johansen, Raymond, Morbidelli, Lambrechts, \& Jacobson}]{BitschEtAl2019}
Bitsch, B., Izidoro, A., Johansen, A., {et~al.} 2019, Astron. Astrophys., 623, A88, \dodoi{10.1051/0004-6361/201834489}

\bibitem[{Bitsch {et~al.}(2015)Bitsch, Lambrechts, \& Johansen}]{BitschEtAl2015}
Bitsch, B., Lambrechts, M., \& Johansen, A. 2015, Astron. Astrophys., 582, A112, \dodoi{10.1051/0004-6361/201526463}

\bibitem[{{Bosman} {et~al.}(2021){Bosman}, {Alarc{\'o}n}, {Bergin}, {Zhang}, {van't Hoff}, {{\"O}berg}, {Guzm{\'a}n}, {Walsh}, {Aikawa}, {Andrews}, {Bergner}, {Booth}, {Cataldi}, {Cleeves}, {Czekala}, {Furuya}, {Huang}, {Ilee}, {Law}, {Le Gal}, {Liu}, {Long}, {Loomis}, {M{\'e}nard}, {Nomura}, {Qi}, {Schwarz}, {Teague}, {Tsukagoshi}, {Yamato}, \& {Wilner}}]{Bosman2021}
{Bosman}, A.~D., {Alarc{\'o}n}, F., {Bergin}, E.~A., {et~al.} 2021, \apjs, 257, 7, \dodoi{10.3847/1538-4365/ac1435}

\bibitem[{Boss(1997)}]{Boss1997}
Boss, A.~P. 1997, Science, 276, 1836, \dodoi{10.1126/science.276.5320.1836}

\bibitem[{Bowler {et~al.}(2020)Bowler, Blunt, \& Nielsen}]{BowlerEtAl2020}
Bowler, B.~P., Blunt, S.~C., \& Nielsen, E.~L. 2020, Astron. J., 159, 63, \dodoi{10.3847/1538-3881/ab5b11}

\bibitem[{Brandt(2021)}]{Brandt2021}
Brandt, T.~D. 2021, Astrophys. J. Suppl. Ser., 254, 42, \dodoi{10.3847/1538-4365/abf93c}

\bibitem[{Brandt {et~al.}(2021)Brandt, Dupuy, Li, Brandt, Zeng, Michalik, Gagliuffi, \& {Raposo-Pulido}}]{BrandtEtAl2021a}
Brandt, T.~D., Dupuy, T.~J., Li, Y., {et~al.} 2021, AJ, 162, 186, \dodoi{10.3847/1538-3881/ac042e}

\bibitem[{{Brown-Sevilla} {et~al.}(2023){Brown-Sevilla}, Maire, Molli{\`e}re, Samland, Feldt, Brandner, Henning, Gratton, Janson, Stolker, Hagelberg, Zurlo, Cantalloube, Boccaletti, Bonnefoy, Chauvin, Desidera, D'Orazi, Lagrange, Langlois, Menard, Mesa, Meyer, Pavlov, Petit, Rochat, Rouan, Schmidt, Vigan, \& Weber}]{Brown-SevillaEtAl2023}
{Brown-Sevilla}, S.~B., Maire, A.~L., Molli{\`e}re, P., {et~al.} 2023, Astron. Astrophys., 673, A98, \dodoi{10.1051/0004-6361/202244826}

\bibitem[{Bryan {et~al.}(2018)Bryan, Benneke, Knutson, Batygin, \& Bowler}]{BryanEtAl2018}
Bryan, M.~L., Benneke, B., Knutson, H.~A., Batygin, K., \& Bowler, B.~P. 2018, Nat. Astron., 2, 138, \dodoi{10.1038/s41550-017-0325-8}

\bibitem[{Bryan {et~al.}(2020{\natexlab{a}})Bryan, Ginzburg, Chiang, Morley, Bowler, Xuan, \& Knutson}]{BryanEtAl2020}
Bryan, M.~L., Ginzburg, S., Chiang, E., {et~al.} 2020{\natexlab{a}}, Astrophys. J., 905, 37, \dodoi{10.3847/1538-4357/abc0ef}

\bibitem[{Bryan {et~al.}(2020{\natexlab{b}})Bryan, Chiang, Bowler, Morley, Millholland, Blunt, Ashok, Nielsen, Ngo, Mawet, \& Knutson}]{BryanEtAl2020a}
Bryan, M.~L., Chiang, E., Bowler, B.~P., {et~al.} 2020{\natexlab{b}}, Astron. J., 159, 181, \dodoi{10.3847/1538-3881/ab76c6}

\bibitem[{Buchner {et~al.}(2014)Buchner, Georgakakis, Nandra, Hsu, Rangel, Brightman, Merloni, Salvato, Donley, \& Kocevski}]{BuchnerEtAl2014}
Buchner, J., Georgakakis, A., Nandra, K., {et~al.} 2014, Astron. Astrophys., 564, A125, \dodoi{10.1051/0004-6361/201322971}

\bibitem[{Burrows {et~al.}(2006)Burrows, Sudarsky, \& Hubeny}]{BurrowsEtAl2006}
Burrows, A., Sudarsky, D., \& Hubeny, I. 2006, Astrophys. J., 640, 1063, \dodoi{10.1086/500293}

\bibitem[{Carvalho \& {Johns-Krull}(2023)}]{CarvalhoJohns-Krull2023}
Carvalho, A., \& {Johns-Krull}, C.~M. 2023, Res. Notes Am. Astron. Soc., 7, 91, \dodoi{10.3847/2515-5172/acd37e}

\bibitem[{Chabrier(2003)}]{Chabrier2003}
Chabrier, G. 2003, Publ. Astron. Soc. Pac., 115, 763, \dodoi{10.1086/376392}

\bibitem[{Chachan {et~al.}(2023)Chachan, Knutson, Lothringer, \& Blake}]{ChachanEtAl2023}
Chachan, Y., Knutson, H.~A., Lothringer, J., \& Blake, G.~A. 2023, Astrophys. J., 943, 112, \dodoi{10.3847/1538-4357/aca614}

\bibitem[{Coles {et~al.}(2019)Coles, Yurchenko, \& Tennyson}]{ColesEtAl2019}
Coles, P.~A., Yurchenko, S.~N., \& Tennyson, J. 2019, Mon. Not. R. Astron. Soc., 490, 4638, \dodoi{10.1093/mnras/stz2778}

\bibitem[{Currie {et~al.}(2023)Currie, Brandt, Brandt, Lacy, Burrows, Guyon, Tamura, Liu, Sagynbayeva, Tobin, Chilcote, Groff, Marois, Thompson, Murphy, Kuzuhara, Lawson, Lozi, Deo, Vievard, Skaf, Uyama, Jovanovic, Martinache, Kasdin, Kudo, McElwain, Janson, Wisniewski, Hodapp, Nishikawa, He{\l}miniak, Kwon, \& Hayashi}]{CurrieEtAl2023}
Currie, T., Brandt, G.~M., Brandt, T.~D., {et~al.} 2023, Science, 380, 198, \dodoi{10.1126/science.abo6192}

\bibitem[{Cushing {et~al.}(2006)Cushing, Roellig, Marley, Saumon, Leggett, Kirkpatrick, Wilson, Sloan, Mainzer, Van~Cleve, \& Houck}]{CushingEtAl2006}
Cushing, M.~C., Roellig, T.~L., Marley, M.~S., {et~al.} 2006, Astrophys. J., 648, 614, \dodoi{10.1086/505637}

\bibitem[{de~Regt {et~al.}(2024)de~Regt, Gandhi, Snellen, Zhang, Christian, \& González~Picos}]{deRegtEtAl2024}
de~Regt, S., Gandhi, S., Snellen, I., {et~al.} 2024

\bibitem[{Delorme {et~al.}(2021)Delorme, Jovanovic, Echeverri, Mawet, Kent~Wallace, Bartos, Cetre, Wizinowich, Ragland, Lilley, Wetherell, Doppmann, Wang, Morris, Ruffio, Martin, Fitzgerald, Ruane, Schofield, Suominen, Calvin, Wang, Magnone, Johnson, Sohn, L{\'o}pez, Bond, Pezzato, Sayson, Chun, \& Skemer}]{DelormeEtAl2021}
Delorme, J.-R., Jovanovic, N., Echeverri, D., {et~al.} 2021, J. Astron. Telesc. Instrum. Syst., 7, 035006, \dodoi{10.1117/1.JATIS.7.3.035006}

\bibitem[{{Do {\'O}} {et~al.}(2023){Do {\'O}}, {O'Neil}, {Konopacky}, {Do}, {Martinez}, {Ruffio}, \& {Ghez}}]{DoOEtAl2023}
{Do {\'O}}, C.~R., {O'Neil}, K.~K., {Konopacky}, Q.~M., {et~al.} 2023, \aj, 166, 48, \dodoi{10.3847/1538-3881/acdc9a}

\bibitem[{D{\"u}rmann \& Kley(2015)}]{DurmannKley2015}
D{\"u}rmann, C., \& Kley, W. 2015, Astron. Astrophys., 574, A52, \dodoi{10.1051/0004-6361/201424837}

\bibitem[{Echeverri {et~al.}(2022)Echeverri, Jovanovic, Delorme, Xin, Schofield, Finnerty, Wang, Xuan, Mawet, Baker, Bartos, Bond, Bryan, Calvin, Cetre, Doppmann, Fitzgerald, Fucik, Horstman, Lopez, Martin, Martin, Mennesson, Morris, Nash, Pezzato, Porter, Ragland, Roberts, Ruane, Ruffio, Sappey, Serabyn, Skemer, Venenciano, Wallace, Wang, \& Wizinowich}]{EcheverriEtAl2022}
Echeverri, D., Jovanovic, N., Delorme, J.-R., {et~al.} 2022, 12184, 121841W, \dodoi{10.1117/12.2630518}

\bibitem[{Erspamer \& North(2003)}]{ErspamerNorth2003}
Erspamer, D., \& North, P. 2003, Astron. Astrophys., 398, 1121, \dodoi{10.1051/0004-6361:20021711}

\bibitem[{Feroz {et~al.}(2009)Feroz, Hobson, \& Bridges}]{FerozEtAl2009}
Feroz, F., Hobson, M.~P., \& Bridges, M. 2009, Mon. Not. R. Astron. Soc., 398, 1601, \dodoi{10.1111/j.1365-2966.2009.14548.x}

\bibitem[{Finnerty {et~al.}(2022)Finnerty, Schofield, Delorme, Sappey, Wang, Ruffio, Mawet, Fitzgerald, Jovanovic, Baker, Bartos, Bond, Bryan, Calvin, Cetre, Doppmann, Echeverri, Lopez, Martin, Morris, Pezzato, Ragland, Ruane, Skemer, Venenciano, Wallace, Wang, Wizinowich, \& Xuan}]{FinnertyEtAl2022}
Finnerty, L., Schofield, T., Delorme, J.-R., {et~al.} 2022, 12184, 121844Y, \dodoi{10.1117/12.2630276}

\bibitem[{{Foreman-Mackey}(2016)}]{Foreman-Mackey2016}
{Foreman-Mackey}, D. 2016, J. Open Source Softw., 1, 24, \dodoi{10.21105/joss.00024}

\bibitem[{Forgan \& Rice(2013)}]{ForganRice2013a}
Forgan, D., \& Rice, K. 2013, Mon. Not. R. Astron. Soc., 432, 3168, \dodoi{10.1093/mnras/stt672}

\bibitem[{{Gaia Collaboration} {et~al.}(2021){Gaia Collaboration}, Brown, Vallenari, Prusti, {de Bruijne}, Babusiaux, Biermann, Creevey, Evans, Eyer, Hutton, Jansen, Jordi, Klioner, Lammers, Lindegren, Luri, Mignard, Panem, Pourbaix, Randich, Sartoretti, Soubiran, Walton, Arenou, {Bailer-Jones}, Bastian, Cropper, Drimmel, Katz, Lattanzi, {van Leeuwen}, Bakker, Cacciari, Casta{\~n}eda, De~Angeli, Ducourant, Fabricius, Fouesneau, Fr{\'e}mat, Guerra, Guerrier, Guiraud, {Jean-Antoine Piccolo}, Masana, Messineo, Mowlavi, Nicolas, Nienartowicz, Pailler, Panuzzo, Riclet, Roux, Seabroke, Sordo, Tanga, Th{\'e}venin, {Gracia-Abril}, Portell, Teyssier, Altmann, Andrae, {Bellas-Velidis}, Benson, Berthier, Blomme, Brugaletta, Burgess, Busso, Carry, Cellino, Cheek, Clementini, Damerdji, Davidson, Delchambre, Dell'Oro, {Fern{\'a}ndez-Hern{\'a}ndez}, Galluccio, {Garc{\'i}a-Lario}, {Garcia-Reinaldos}, {Gonz{\'a}lez-N{\'u}{\~n}ez}, Gosset, Haigron, Halbwachs, Hambly, Harrison, Hatzidimitriou, Heiter, Hern{\'a}ndez, Hestroffer, Hodgkin, Holl, Jan{\ss}en, {Jevardat de Fombelle}, Jordan, {Krone-Martins}, Lanzafame, L{\"o}ffler, Lorca, Manteiga, Marchal, Marrese, Moitinho, Mora, Muinonen, Osborne, Pancino, Pauwels, Petit, {Recio-Blanco}, Richards, Riello, Rimoldini, Robin, Roegiers, Rybizki, Sarro, Siopis, Smith, Sozzetti, Ulla, Utrilla, {van Leeuwen}, {van Reeven}, Abbas, Abreu~Aramburu, Accart, Aerts, Aguado, Ajaj, Altavilla, {\'A}lvarez, {\'A}lvarez Cid-Fuentes, Alves, Anderson, Anglada~Varela, Antoja, Audard, Baines, Baker, {Balaguer-N{\'u}{\~n}ez}, Balbinot, Balog, Barache, Barbato, Barros, Barstow, Bartolom{\'e}, Bassilana, Bauchet, {Baudesson-Stella}, Becciani, Bellazzini, Bernet, Bertone, Bianchi, {Blanco-Cuaresma}, Boch, Bombrun, Bossini, Bouquillon, Bragaglia, Bramante, Breedt, Bressan, Brouillet, Bucciarelli, Burlacu, Busonero, Butkevich, Buzzi, Caffau, Cancelliere, C{\'a}novas, {Cantat-Gaudin}, Carballo, Carlucci, Carnerero, Carrasco, Casamiquela, Castellani, {Castro-Ginard}, Castro~Sampol, Chaoul, Charlot, Chemin, Chiavassa, Cioni, Comoretto, Cooper, Cornez, Cowell, Crifo, Crosta, Crowley, Dafonte, Dapergolas, David, David, {de Laverny}, De~Luise, De~March, De~Ridder, {de Souza}, {de Teodoro}, {de Torres}, {del Peloso}, {del Pozo}, Delbo, Delgado, Delgado, Delisle, Di~Matteo, Diakite, Diener, Distefano, Dolding, Eappachen, Edvardsson, Enke, Esquej, Fabre, Fabrizio, Faigler, Fedorets, Fernique, Fienga, Figueras, Fouron, Fragkoudi, Fraile, Franke, Gai, Garabato, {Garcia-Gutierrez}, {Garc{\'i}a-Torres}, Garofalo, Gavras, Gerlach, Geyer, Giacobbe, Gilmore, Girona, Giuffrida, Gomel, Gomez, {Gonzalez-Santamaria}, {Gonz{\'a}lez-Vidal}, Granvik, {Guti{\'e}rrez-S{\'a}nchez}, Guy, Hauser, Haywood, Helmi, Hidalgo, Hilger, H{\l}adczuk, Hobbs, Holland, Huckle, Jasniewicz, Jonker, Juaristi~Campillo, Julbe, Karbevska, Kervella, Khanna, Kochoska, Kontizas, Kordopatis, Korn, {Kostrzewa-Rutkowska}, Kruszy{\'n}ska, Lambert, Lanza, Lasne, Le~Campion, Le~Fustec, Lebreton, Lebzelter, Leccia, Leclerc, {Lecoeur-Taibi}, Liao, Licata, Lindstr{\o}m, Lister, Livanou, Lobel, Madrero~Pardo, Managau, Mann, Marchant, Marconi, Marcos~Santos, Marinoni, Marocco, Marshall, Martin~Polo, {Mart{\'i}n-Fleitas}, Masip, Massari, {Mastrobuono-Battisti}, Mazeh, McMillan, Messina, Michalik, Millar, Mints, Molina, Molinaro, Moln{\'a}r, Montegriffo, Mor, Morbidelli, Morel, Morris, Mulone, Munoz, Muraveva, Murphy, Musella, Noval, Ord{\'e}novic, Orr{\`u}, Osinde, Pagani, Pagano, Palaversa, Palicio, Panahi, Pawlak, Pe{\~n}alosa~Esteller, Penttil{\"a}, Piersimoni, Pineau, Plachy, Plum, Poggio, Poretti, Poujoulet, Pr{\v s}a, Pulone, Racero, Ragaini, Rainer, Raiteri, Rambaux, Ramos, {Ramos-Lerate}, Re~Fiorentin, Regibo, Reyl{\'e}, Ripepi, Riva, Rixon, Robichon, Robin, Roelens, Rohrbasser, {Romero-G{\'o}mez}, Rowell, Royer, Rybicki, Sadowski, Sagrist{\`a}~Sell{\'e}s, Sahlmann, Salgado, Salguero, Samaras, Sanchez~Gimenez, Sanna, Santove{\~n}a, Sarasso, Schultheis, Sciacca, Segol, Segovia, S{\'e}gransan, Semeux, Shahaf, Siddiqui, Siebert, Siltala, Slezak, Smart, Solano, Solitro, Souami, Souchay, Spagna, Spoto, Steele, Steidelm{\"u}ller, Stephenson, S{\"u}veges, Szabados, {Szegedi-Elek}, Taris, Tauran, Taylor, Teixeira, Thuillot, Tonello, Torra, Torra, Turon, Unger, Vaillant, {van Dillen}, Vanel, Vecchiato, Viala, Vicente, Voutsinas, Weiler, Wevers, Wyrzykowski, Yoldas, Yvard, Zhao, Zorec, Zucker, Zurbach, \& Zwitter}]{GaiaCollaborationEtAl2021}
{Gaia Collaboration}, Brown, A. G.~A., Vallenari, A., {et~al.} 2021, Astron. Astrophys., 649, A1, \dodoi{10.1051/0004-6361/202039657}

\bibitem[{Ginzburg \& Chiang(2020)}]{GinzburgChiang2020}
Ginzburg, S., \& Chiang, E. 2020, Mon. Not. R. Astron. Soc., 491, L34, \dodoi{10.1093/mnrasl/slz164}

\bibitem[{{GRAVITY Collaboration} {et~al.}(2020){GRAVITY Collaboration}, Nowak, Lacour, Molli{\`e}re, Wang, Charnay, {van Dishoeck}, Abuter, Amorim, Berger, Beust, Bonnefoy, Bonnet, Brandner, Buron, Cantalloube, Collin, Chapron, Cl{\'e}net, Coud{\'e} Du~Foresto, {de Zeeuw}, Dembet, Dexter, Duvert, Eckart, Eisenhauer, F{\"o}rster~Schreiber, F{\'e}dou, Garcia~Lopez, Gao, Gendron, Genzel, Gillessen, Hau{\ss}mann, Henning, Hippler, Hubert, Jocou, Kervella, Lagrange, Lapeyr{\`e}re, Le~Bouquin, L{\'e}na, Maire, Ott, Paumard, Paladini, Perraut, Perrin, Pueyo, Pfuhl, Rabien, Rau, {Rodr{\'i}guez-Coira}, Rousset, Scheithauer, Shangguan, Straub, Straubmeier, Sturm, Tacconi, Vincent, Widmann, Wieprecht, Wiezorrek, Woillez, Yazici, \& Ziegler}]{GRAVITYCollaborationEtAl2020}
{GRAVITY Collaboration}, Nowak, M., Lacour, S., {et~al.} 2020, Astron. Astrophys., 633, A110, \dodoi{10.1051/0004-6361/201936898}

\bibitem[{Hargreaves {et~al.}(2020)Hargreaves, Gordon, Rey, Nikitin, Tyuterev, Kochanov, \& Rothman}]{HargreavesEtAl2020}
Hargreaves, R.~J., Gordon, I.~E., Rey, M., {et~al.} 2020, Astrophys. J. Suppl. Ser., 247, 55, \dodoi{10.3847/1538-4365/ab7a1a}

\bibitem[{Harris {et~al.}(2020)Harris, Millman, van~der Walt, Gommers, Virtanen, Cournapeau, Wieser, Taylor, Berg, Smith, Kern, Picus, Hoyer, van Kerkwijk, Brett, Haldane, del R{\'{i}}o, Wiebe, Peterson, G{\'{e}}rard-Marchant, Sheppard, Reddy, Weckesser, Abbasi, Gohlke, \& Oliphant}]{Harris2020}
Harris, C.~R., Millman, K.~J., van~der Walt, S.~J., {et~al.} 2020, Nature, 585, 357, \dodoi{10.1038/s41586-020-2649-2}

\bibitem[{Hinkel {et~al.}(2014)Hinkel, Timmes, Young, Pagano, \& Turnbull}]{HinkelEtAl2014}
Hinkel, N.~R., Timmes, F.~X., Young, P.~A., Pagano, M.~D., \& Turnbull, M.~C. 2014, Astron. J., 148, 54, \dodoi{10.1088/0004-6256/148/3/54}

\bibitem[{Hoch {et~al.}(2023)Hoch, Konopacky, Theissen, Ruffio, Barman, Rickman, Perrin, Macintosh, \& Marois}]{HochEtAl2023}
Hoch, K. K.~W., Konopacky, Q.~M., Theissen, C.~A., {et~al.} 2023, Astron. J., 166, 85, \dodoi{10.3847/1538-3881/ace442}

\bibitem[{Hoeijmakers {et~al.}(2018)Hoeijmakers, Schwarz, Snellen, {de Kok}, Bonnefoy, Chauvin, Lagrange, \& Girard}]{HoeijmakersEtAl2018}
Hoeijmakers, H.~J., Schwarz, H., Snellen, I. A.~G., {et~al.} 2018, Astron. Astrophys., 617, A144, \dodoi{10.1051/0004-6361/201832902}

\bibitem[{Hsu {et~al.}(2021)Hsu, Burgasser, Theissen, Gelino, Birky, Diamant, Bardalez~Gagliuffi, Aganze, Blake, \& Faherty}]{HsuEtAl2021}
Hsu, C.-C., Burgasser, A.~J., Theissen, C.~A., {et~al.} 2021, Astrophys. J. Suppl. Ser., 257, 45, \dodoi{10.3847/1538-4365/ac1c7d}

\bibitem[{Hunter(2007)}]{Hunter2007}
Hunter, J.~D. 2007, Computing in Science \& Engineering, 9, 90, \dodoi{10.1109/MCSE.2007.55}

\bibitem[{Husser {et~al.}(2013)Husser, {Wende-von Berg}, Dreizler, Homeier, Reiners, Barman, \& Hauschildt}]{HusserEtAl2013}
Husser, T.~O., {Wende-von Berg}, S., Dreizler, S., {et~al.} 2013, Astron. Astrophys., 553, A6, \dodoi{10.1051/0004-6361/201219058}

\bibitem[{Inglis {et~al.}(2024)Inglis, Wallack, Xuan, Knutson, Chachan, Bryan, Bowler, Iyer, Kataria, \& Benneke}]{InglisEtAl2024}
Inglis, J., Wallack, N.~L., Xuan, J.~W., {et~al.} 2024, Atmospheric Retrievals of the Young Giant Planet ROXs 42B b from Low- and High-Resolution Spectroscopy,  arXiv.
\newblock \doeprint{2402.09533}

\bibitem[{{Kempton} \& {Knutson}(2024)}]{KemptonKnutson2024}
{Kempton}, E. M.~R., \& {Knutson}, H.~A. 2024, arXiv e-prints, arXiv:2404.15430, \dodoi{10.48550/arXiv.2404.15430}

\bibitem[{Konopacky {et~al.}(2013)Konopacky, Barman, Macintosh, \& Marois}]{KonopackyEtAl2013}
Konopacky, Q.~M., Barman, T.~S., Macintosh, B.~A., \& Marois, C. 2013, Science, 339, 1398, \dodoi{10.1126/science.1232003}

\bibitem[{Kotani {et~al.}(2020)Kotani, Kawahara, Ishizuka, Jovanovic, Vievard, Lozi, Sahoo, Guyon, Yoneta, \& Tamura}]{KotaniEtAl2020a}
Kotani, T., Kawahara, H., Ishizuka, M., {et~al.} 2020, 11448, 1144878, \dodoi{10.1117/12.2561755}

\bibitem[{Kratter \& Lodato(2016)}]{KratterLodato2016}
Kratter, K., \& Lodato, G. 2016, Annu. Rev. Astron. Astrophys., 54, 271, \dodoi{10.1146/annurev-astro-081915-023307}

\bibitem[{Lambrechts \& Johansen(2012)}]{LambrechtsJohansen2012}
Lambrechts, M., \& Johansen, A. 2012, Astron. Astrophys., 544, A32, \dodoi{10.1051/0004-6361/201219127}

\bibitem[{Landman {et~al.}(2024)Landman, Stolker, Snellen, Costes, {de Regt}, Zhang, Gandhi, Molliere, Kesseli, Vigan, \& {Sanchez-L{\'o}pez}}]{LandmanEtAl2024}
Landman, R., Stolker, T., Snellen, I. A.~G., {et~al.} 2024, Astron. Astrophys., 682, A48, \dodoi{10.1051/0004-6361/202347846}

\bibitem[{{Le Gal} {et~al.}(2021){Le Gal}, {{\"O}berg}, {Teague}, {Loomis}, {Law}, {Walsh}, {Bergin}, {M{\'e}nard}, {Wilner}, {Andrews}, {Aikawa}, {Booth}, {Cataldi}, {Bergner}, {Bosman}, {Cleeves}, {Czekala}, {Furuya}, {Guzm{\'a}n}, {Huang}, {Ilee}, {Nomura}, {Qi}, {Schwarz}, {Tsukagoshi}, {Yamato}, \& {Zhang}}]{LeGal2021}
{Le Gal}, R., {{\"O}berg}, K.~I., {Teague}, R., {et~al.} 2021, \apjs, 257, 12, \dodoi{10.3847/1538-4365/ac2583}

\bibitem[{Li {et~al.}(2015)Li, Gordon, Rothman, Tan, Hu, Kassi, Campargue, \& Medvedev}]{LiEtAl2015}
Li, G., Gordon, I.~E., Rothman, L.~S., {et~al.} 2015, Astrophys. J. Suppl. Ser., 216, 15, \dodoi{10.1088/0067-0049/216/1/15}

\bibitem[{Lothringer {et~al.}(2021)Lothringer, Rustamkulov, Sing, Gibson, Wilson, \& Schlaufman}]{LothringerEtAl2021}
Lothringer, J.~D., Rustamkulov, Z., Sing, D.~K., {et~al.} 2021, Astrophys. J., 914, 12, \dodoi{10.3847/1538-4357/abf8a9}

\bibitem[{Madhusudhan(2012)}]{Madhusudhan2012}
Madhusudhan, N. 2012, Astrophys. J., 758, 36, \dodoi{10.1088/0004-637X/758/1/36}

\bibitem[{Madhusudhan {et~al.}(2014)Madhusudhan, Amin, \& Kennedy}]{MadhusudhanEtAl2014}
Madhusudhan, N., Amin, M.~A., \& Kennedy, G.~M. 2014, Astrophys. J., 794, L12, \dodoi{10.1088/2041-8205/794/1/L12}

\bibitem[{Marley {et~al.}(2021)Marley, Saumon, Visscher, Lupu, Freedman, Morley, Fortney, Seay, Smith, Teal, \& Wang}]{MarleyEtAl2021}
Marley, M.~S., Saumon, D., Visscher, C., {et~al.} 2021, Astrophys. J., 920, 85, \dodoi{10.3847/1538-4357/ac141d}

\bibitem[{Martin {et~al.}(2018)Martin, Fitzgerald, McLean, Doppmann, Kassis, Aliado, Canfield, Johnson, Kress, Lanclos, Magnone, Sohn, Wang, \& Weiss}]{MartinEtAl2018}
Martin, E.~C., Fitzgerald, M.~P., McLean, I.~S., {et~al.} 2018, 10702, 107020A, \dodoi{10.1117/12.2312266}

\bibitem[{Mawet {et~al.}(2017)Mawet, Ruane, Xuan, Echeverri, Klimovich, Randolph, Fucik, Wallace, Wang, Vasisht, Dekany, Mennesson, Choquet, Delorme, \& Serabyn}]{MawetEtAl2017}
Mawet, D., Ruane, G., Xuan, W., {et~al.} 2017, Astrophys. J., 838, 92, \dodoi{10.3847/1538-4357/aa647f}

\bibitem[{McLean {et~al.}(1998)McLean, Becklin, Bendiksen, Brims, Canfield, Figer, Graham, Hare, Lacayanga, Larkin, Larson, Levenson, Magnone, Teplitz, \& Wong}]{McLeanEtAl1998}
McLean, I.~S., Becklin, E.~E., Bendiksen, O., {et~al.} 1998, 3354, 566, \dodoi{10.1117/12.317283}

\bibitem[{Molli{\`e}re {et~al.}(2017)Molli{\`e}re, {van Boekel}, Bouwman, Henning, Lagage, \& Min}]{MolliereEtAl2017}
Molli{\`e}re, P., {van Boekel}, R., Bouwman, J., {et~al.} 2017, Astron. Astrophys., 600, A10, \dodoi{10.1051/0004-6361/201629800}

\bibitem[{Molli{\`e}re {et~al.}(2019)Molli{\`e}re, Wardenier, {van Boekel}, Henning, Molaverdikhani, \& Snellen}]{MolliereEtAl2019}
Molli{\`e}re, P., Wardenier, J.~P., {van Boekel}, R., {et~al.} 2019, Astron. Astrophys., 627, A67, \dodoi{10.1051/0004-6361/201935470}

\bibitem[{Molli{\`e}re {et~al.}(2020)Molli{\`e}re, Stolker, Lacour, Otten, Shangguan, Charnay, Molyarova, Nowak, Henning, Marleau, Semenov, {van Dishoeck}, Eisenhauer, Garcia, Garcia~Lopez, Girard, Greenbaum, Hinkley, Kervella, Kreidberg, Maire, Nasedkin, Pueyo, Snellen, Vigan, Wang, {de Zeeuw}, \& Zurlo}]{MolliereEtAl2020}
Molli{\`e}re, P., Stolker, T., Lacour, S., {et~al.} 2020, Astron. Astrophys., 640, A131, \dodoi{10.1051/0004-6361/202038325}

\bibitem[{Molli{\`e}re {et~al.}(2022)Molli{\`e}re, Molyarova, Bitsch, Henning, Schneider, Kreidberg, Eistrup, Burn, Nasedkin, Semenov, Mordasini, Schlecker, Schwarz, Lacour, Nowak, \& Schulik}]{MolliereEtAl2022a}
Molli{\`e}re, P., Molyarova, T., Bitsch, B., {et~al.} 2022, Astrophys. J., 934, 74, \dodoi{10.3847/1538-4357/ac6a56}

\bibitem[{Mordasini {et~al.}(2016)Mordasini, {van Boekel}, Molli{\`e}re, Henning, \& Benneke}]{MordasiniEtAl2016}
Mordasini, C., {van Boekel}, R., Molli{\`e}re, P., Henning, {\relax Th}., \& Benneke, B. 2016, Astrophys. J., 832, 41, \dodoi{10.3847/0004-637X/832/1/41}

\bibitem[{Murphy \& Paunzen(2017)}]{MurphyPaunzen2017}
Murphy, S.~J., \& Paunzen, E. 2017, Mon. Not. R. Astron. Soc., 466, 546, \dodoi{10.1093/mnras/stw3141}

\bibitem[{{Nagpal} {et~al.}(2023){Nagpal}, {Blunt}, {Bowler}, {Dupuy}, {Nielsen}, \& {Wang}}]{NagpalEtAl2023}
{Nagpal}, V., {Blunt}, S., {Bowler}, B.~P., {et~al.} 2023, \aj, 165, 32, \dodoi{10.3847/1538-3881/ac9fd2}

\bibitem[{Nasedkin {et~al.}(2024)Nasedkin, Molli{\`e}re, Lacour, Nowak, Kreidberg, Stolker, Wang, Balmer, Kammerer, Shangguan, Abuter, Amorim, {Asensio-Torres}, Benisty, Berger, Beust, Blunt, Boccaletti, Bonnefoy, Bonnet, Bordoni, Bourdarot, Brandner, Cantalloube, Caselli, Charnay, Chauvin, Chavez, Choquet, Christiaens, Cl{\'e}net, du~Foresto, Cridland, Davies, Dembet, Dexter, Drescher, Duvert, Eckart, Eisenhauer, Schreiber, Garcia, Lopez, Gendron, Genzel, Gillessen, Girard, Grant, Haubois, Hei{\ss}el, Henning, Hinkley, Hippler, Houll{\'e}, Hubert, Jocou, Keppler, Kervella, Kurtovic, Lagrange, Lapeyr{\`e}re, Bouquin, Lutz, Maire, Mang, Marleau, M{\'e}rand, Monnier, Mordasini, Ott, Otten, Paladini, Paumard, Perraut, Perrin, Pfuhl, Pourr{\'e}, Pueyo, Ribeiro, Rickman, Ruffio, Rustamkulov, Shimizu, Sing, Stadler, Straub, Straubmeier, Sturm, Tacconi, {van Dishoeck}, Vigan, Vincent, {von Fellenberg}, Widmann, Winterhalder, Woillez, Yazici, \& Collaboration}]{NasedkinEtAl2024}
Nasedkin, E., Molli{\`e}re, P., Lacour, S., {et~al.} 2024, Four-of-a-kind? Comprehensive atmospheric characterisation of the HR 8799 planets with VLTI/GRAVITY,  arXiv.
\newblock \doeprint{2404.03776}

\bibitem[{Nielsen {et~al.}(2019)Nielsen, De~Rosa, Macintosh, Wang, Ruffio, Chiang, Marley, Saumon, Savransky, Ammons, Bailey, Barman, Blain, Bulger, Burrows, Chilcote, Cotten, Czekala, Doyon, Duch{\^e}ne, Esposito, Fabrycky, Fitzgerald, Follette, Fortney, Gerard, Goodsell, Graham, Greenbaum, Hibon, Hinkley, Hirsch, Hom, Hung, Dawson, Ingraham, Kalas, Konopacky, Larkin, Lee, Lin, Maire, Marchis, Marois, Metchev, {Millar-Blanchaer}, Morzinski, Oppenheimer, Palmer, Patience, Perrin, Poyneer, Pueyo, Rafikov, Rajan, Rameau, Rantakyr{\"o}, Ren, Schneider, Sivaramakrishnan, Song, Soummer, Tallis, Thomas, {Ward-Duong}, \& Wolff}]{NielsenEtAl2019}
Nielsen, E.~L., De~Rosa, R.~J., Macintosh, B., {et~al.} 2019, Astron. J., 158, 13, \dodoi{10.3847/1538-3881/ab16e9}

\bibitem[{{\"O}berg {et~al.}(2011){\"O}berg, {Murray-Clay}, \& Bergin}]{ObergEtAl2011}
{\"O}berg, K.~I., {Murray-Clay}, R., \& Bergin, E.~A. 2011, Astrophys. J., 743, L16, \dodoi{10.1088/2041-8205/743/1/L16}

\bibitem[{{{\"O}berg} {et~al.}(2021){{\"O}berg}, {Guzm{\'a}n}, {Walsh}, {Aikawa}, {Bergin}, {Law}, {Loomis}, {Alarc{\'o}n}, {Andrews}, {Bae}, {Bergner}, {Boehler}, {Booth}, {Bosman}, {Calahan}, {Cataldi}, {Cleeves}, {Czekala}, {Furuya}, {Huang}, {Ilee}, {Kurtovic}, {Le Gal}, {Liu}, {Long}, {M{\'e}nard}, {Nomura}, {P{\'e}rez}, {Qi}, {Schwarz}, {Sierra}, {Teague}, {Tsukagoshi}, {Yamato}, {van't Hoff}, {Waggoner}, {Wilner}, \& {Zhang}}]{Oberg2021}
{{\"O}berg}, K.~I., {Guzm{\'a}n}, V.~V., {Walsh}, C., {et~al.} 2021, \apjs, 257, 1, \dodoi{10.3847/1538-4365/ac1432}

\bibitem[{{Offner} {et~al.}(2023){Offner}, {Moe}, {Kratter}, {Sadavoy}, {Jensen}, \& {Tobin}}]{OffnerEtAl2023}
{Offner}, S.~S.~R., {Moe}, M., {Kratter}, K.~M., {et~al.} 2023, in Astronomical Society of the Pacific Conference Series, Vol. 534, Protostars and Planets VII, ed. S.~{Inutsuka}, Y.~{Aikawa}, T.~{Muto}, K.~{Tomida}, \& M.~{Tamura}, 275, \dodoi{10.48550/arXiv.2203.10066}

\bibitem[{{Palma-Bifani} {et~al.}(2023){Palma-Bifani}, {Chauvin}, {Bonnefoy}, {Rojo}, {Petrus}, {Rodet}, {Langlois}, {Allard}, {Charnay}, {Desgrange}, {Homeier}, {Lagrange}, {Beuzit}, {Baudoz}, {Boccaletti}, {Chomez}, {Delorme}, {Desidera}, {Feldt}, {Ginski}, {Gratton}, {Maire}, {Meyer}, {Samland}, {Snellen}, {Vigan}, \& {Zhang}}]{Palma-BifaniEtAl2023}
{Palma-Bifani}, P., {Chauvin}, G., {Bonnefoy}, M., {et~al.} 2023, \aap, 670, A90, \dodoi{10.1051/0004-6361/202244294}

\bibitem[{{Palma-Bifani} {et~al.}(2024){Palma-Bifani}, Chauvin, Borja, Bonnefoy, Petrus, Mesa, De~Rosa, Gratton, Baudoz, Boccaletti, Charnay, Desgrange, Tremblin, \& Vigan}]{Palma-BifaniEtAl2024}
{Palma-Bifani}, P., Chauvin, G., Borja, D., {et~al.} 2024, Atmospheric properties of AF Lep b with forward modeling,  arXiv, \dodoi{10.48550/arXiv.2401.05491}

\bibitem[{{Petrus} {et~al.}(2024){Petrus}, {Whiteford}, {Patapis}, {Biller}, {Skemer}, {Hinkley}, {Su{\'a}rez}, {Palma-Bifani}, {Morley}, {Tremblin}, {Charnay}, {Vos}, {Wang}, {Stone}, {Bonnefoy}, {Chauvin}, {Miles}, {Carter}, {Lueber}, {Helling}, {Sutlieff}, {Janson}, {Gonzales}, {Hoch}, {Absil}, {Balmer}, {Boccaletti}, {Bonavita}, {Booth}, {Bowler}, {Briesemeister}, {Bryan}, {Calissendorff}, {Cantalloube}, {Chen}, {Choquet}, {Christiaens}, {Cugno}, {Currie}, {Danielski}, {De Furio}, {Dupuy}, {Factor}, {Faherty}, {Fitzgerald}, {Fortney}, {Franson}, {Girard}, {Grady}, {Henning}, {Hines}, {Hood}, {Howe}, {Kalas}, {Kammerer}, {Kennedy}, {Kenworthy}, {Kervella}, {Kim}, {Kitzmann}, {Kraus}, {Kuzuhara}, {Lagage}, {Lagrange}, {Lawson}, {Lazzoni}, {Leisenring}, {Lew}, {Liu}, {Liu}, {Llop-Sayson}, {Lloyd}, {Macintosh}, {M{\^a}lin}, {Manjavacas}, {Marino}, {Marley}, {Marois}, {Martinez}, {Matthews}, {Matthews}, {Mawet}, {Mazoyer}, {McElwain}, {Metchev}, {Meyer}, {Millar-Blanchaer}, {Molli{\`e}re}, {Moran}, {Mukherjee}, {Pantin}, {Perrin}, {Pueyo}, {Quanz}, {Quirrenbach}, {Ray}, {Rebollido}, {Adams Redai}, {Ren}, {Rickman}, {Sallum}, {Samland}, {Sargent}, {Schlieder}, {Stapelfeldt}, {Tamura}, {Tan}, {Theissen}, {Uyama}, {Vasist}, {Vigan}, {Wagner}, {Ward-Duong}, {Wolff}, {Worthen}, {Wyatt}, {Ygouf}, {Zurlo}, {Zhang}, {Zhang}, {Zhang}, \& {Zhou}}]{Petrus2024}
{Petrus}, S., {Whiteford}, N., {Patapis}, P., {et~al.} 2024, \apjl, 966, L11, \dodoi{10.3847/2041-8213/ad3e7c}

\bibitem[{Pollack {et~al.}(1996)Pollack, Hubickyj, Bodenheimer, Lissauer, Podolak, \& Greenzweig}]{PollackEtAl1996}
Pollack, J.~B., Hubickyj, O., Bodenheimer, P., {et~al.} 1996, Icarus, 124, 62, \dodoi{10.1006/icar.1996.0190}

\bibitem[{Polyansky {et~al.}(2018)Polyansky, Kyuberis, Zobov, Tennyson, Yurchenko, \& Lodi}]{PolyanskyEtAl2018}
Polyansky, O.~L., Kyuberis, A.~A., Zobov, N.~F., {et~al.} 2018, Mon. Not. R. Astron. Soc., 480, 2597, \dodoi{10.1093/mnras/sty1877}

\bibitem[{Qi {et~al.}(2015)Qi, {\"O}berg, Andrews, Wilner, Bergin, Hughes, Hogherheijde, \& D'Alessio}]{QiEtAl2015}
Qi, C., {\"O}berg, K.~I., Andrews, S.~M., {et~al.} 2015, Astrophys. J., 813, 128, \dodoi{10.1088/0004-637X/813/2/128}

\bibitem[{Robert {et~al.}(2018)Robert, Crida, Lega, M{\'e}heut, \& Morbidelli}]{RobertEtAl2018}
Robert, C. M.~T., Crida, A., Lega, E., M{\'e}heut, H., \& Morbidelli, A. 2018, A\&A, 617, A98, \dodoi{10.1051/0004-6361/201833539}

\bibitem[{Rothman {et~al.}(2010)Rothman, Gordon, Barber, Dothe, Gamache, Goldman, Perevalov, Tashkun, \& Tennyson}]{RothmanEtAl2010}
Rothman, L.~S., Gordon, I.~E., Barber, R.~J., {et~al.} 2010, J. Quant. Spectrosc. Radiat. Transf., 111, 2139, \dodoi{10.1016/j.jqsrt.2010.05.001}

\bibitem[{Ruffio {et~al.}(2019)Ruffio, Macintosh, Konopacky, Barman, De~Rosa, Wang, Wilcomb, Czekala, \& Marois}]{RuffioEtAl2019}
Ruffio, J.-B., Macintosh, B., Konopacky, Q.~M., {et~al.} 2019, Astron. J., 158, 200, \dodoi{10.3847/1538-3881/ab4594}

\bibitem[{Ruffio {et~al.}(2021)Ruffio, Konopacky, Barman, Macintosh, Wilcomb, De~Rosa, Wang, Czekala, \& Marois}]{RuffioEtAl2021}
Ruffio, J.-B., Konopacky, Q.~M., Barman, T., {et~al.} 2021, AJ, 162, 290, \dodoi{10.3847/1538-3881/ac273a}

\bibitem[{Ruffio {et~al.}(2023)Ruffio, Perrin, Hoch, Kammerer, Konopacky, Pueyo, Rickman, Theissen, Agrawal, Greenbaum, Miles, Barman, Balmer, {Llop-Sayson}, Girard, Rebollido, Soummer, Allen, Anderson, Beichman, Bellini, Bryden, Espinoza, Glidden, Huang, Lewis, Libralato, Louie, Sohn, Seager, {van der Marel}, Wakeford, Watkins, Ygouf, \& Mountai}]{RuffioEtAl2023}
Ruffio, J.-B., Perrin, M.~D., Hoch, K. K.~W., {et~al.} 2023, JWST-TST High Contrast: Achieving direct spectroscopy of faint substellar companions next to bright stars with the NIRSpec IFU,  arXiv, \dodoi{10.48550/arXiv.2310.09902}

\bibitem[{Schib {et~al.}(2021)Schib, Mordasini, Wenger, Marleau, \& Helled}]{SchibEtAl2021}
Schib, O., Mordasini, C., Wenger, N., Marleau, G.~D., \& Helled, R. 2021, Astron. Astrophys., 645, A43, \dodoi{10.1051/0004-6361/202039154}

\bibitem[{Schwarz {et~al.}(2016)Schwarz, Ginski, {de Kok}, Snellen, Brogi, \& Birkby}]{SchwarzEtAl2016}
Schwarz, H., Ginski, C., {de Kok}, R.~J., {et~al.} 2016, Astron. Astrophys., 593, A74, \dodoi{10.1051/0004-6361/201628908}

\bibitem[{Snellen {et~al.}(2015)Snellen, {de Kok}, Birkby, Brandl, Brogi, Keller, Kenworthy, Schwarz, \& Stuik}]{SnellenEtAl2015}
Snellen, I., {de Kok}, R., Birkby, J.~L., {et~al.} 2015, Astron. Astrophys., 576, A59, \dodoi{10.1051/0004-6361/201425018}

\bibitem[{Spiegel \& Burrows(2012)}]{SpiegelBurrows2012}
Spiegel, D.~S., \& Burrows, A. 2012, Astrophys. J., 745, 174, \dodoi{10.1088/0004-637X/745/2/174}

\bibitem[{Turrini {et~al.}(2021)Turrini, Schisano, Fonte, Molinari, Politi, Fedele, Pani{\'c}, Kama, Changeat, \& Tinetti}]{TurriniEtAl2021}
Turrini, D., Schisano, E., Fonte, S., {et~al.} 2021, Astrophys. J., 909, 40, \dodoi{10.3847/1538-4357/abd6e5}

\bibitem[{Veras {et~al.}(2009)Veras, Crepp, \& Ford}]{VerasEtAl2009}
Veras, D., Crepp, J.~R., \& Ford, E.~B. 2009, Astrophys. J., 696, 1600, \dodoi{10.1088/0004-637X/696/2/1600}

\bibitem[{Vigan {et~al.}(2023)Vigan, Morsy, Lopez, Otten, Garcia, Costes, Muslimov, Viret, Charles, Zins, Murray, Costille, Paufique, Seemann, Houll{\'e}, {Anwand-Heerwart}, Phillips, Abinanti, Balard, Baraffe, Benedetti, Blanchard, Blanco, Beuzit, Choquet, Cristofari, Desidera, Dohlen, Dorn, Ely, Fuenteseca, Garcia, Jaquet, Jaubert, Kasper, Merrer, Maire, N'Diaye, Pallanca, Popovic, Pourcelot, Reiners, Rochat, Sehim, Schmutzer, Smette, Tchoubaklian, Tomlinson, \& Soto}]{ViganEtAl2023}
Vigan, A., Morsy, M.~E., Lopez, M., {et~al.} 2023, First light of VLT/HiRISE: High-resolution spectroscopy of young giant exoplanets,  arXiv, \dodoi{10.48550/arXiv.2309.12390}

\bibitem[{Virtanen {et~al.}(2020)Virtanen, Gommers, Oliphant, Haberland, Reddy, Cournapeau, Burovski, Peterson, Weckesser, Bright, {van der Walt}, Brett, Wilson, Millman, Mayorov, Nelson, Jones, Kern, Larson, Carey, Polat, Feng, Moore, VanderPlas, Laxalde, Perktold, Cimrman, Henriksen, Quintero, Harris, Archibald, Ribeiro, Pedregosa, {van Mulbregt}, \& {SciPy 1. 0 Contributors}}]{VirtanenEtAl2020}
Virtanen, P., Gommers, R., Oliphant, T.~E., {et~al.} 2020, Nat. Methods, 17, 261, \dodoi{10.1038/s41592-019-0686-2}

\bibitem[{Vos {et~al.}(2020)Vos, Biller, Allers, Faherty, Liu, Metchev, Eriksson, Manjavacas, Dupuy, Janson, {Radigan-Hoffman}, Crossfield, Bonnefoy, Best, Homeier, Schlieder, Brandner, Henning, Bonavita, \& Buenzli}]{VosEtAl2020}
Vos, J.~M., Biller, B.~A., Allers, K.~N., {et~al.} 2020, Astron. J., 160, 38, \dodoi{10.3847/1538-3881/ab9642}

\bibitem[{Wang {et~al.}(2022)Wang, Wang, Ruffio, Blake, Mawet, Baker, Bartos, Bond, Calvin, Cetre, Delorme, Doppmann, Echeverri, Finnerty, Fitzgerald, Jovanovic, Lopez, Martin, Morris, Pezzato, Ragland, Ruane, Sappey, Schofield, Skemer, Venenciano, Wallace, Wizinowich, Xuan, Bryan, Roy, \& Wallack}]{WangEtAl2022}
Wang, J., Wang, J.~J., Ruffio, J.-B., {et~al.} 2022, AJ, 165, 4, \dodoi{10.3847/1538-3881/ac9f19}

\bibitem[{Wang {et~al.}(2021)Wang, Ruffio, Morris, Delorme, Jovanovic, Pezzato, Echeverri, Finnerty, Hood, Zanazzi, Bryan, Bond, Cetre, Martin, Mawet, Skemer, Baker, Xuan, Wallace, Wang, Bartos, Blake, Boden, Buzard, Calvin, Chun, Doppmann, Dupuy, Duch{\^e}ne, Feng, Fitzgerald, Fortney, Freedman, Knutson, Konopacky, Lilley, Liu, Lopez, Lupu, Marley, Meshkat, Miles, {Millar-Blanchaer}, Ragland, Roy, Ruane, Sappey, Schofield, Weiss, Wetherell, Wizinowich, \& Ygouf}]{WangEtAl2021}
Wang, J.~J., Ruffio, J.-B., Morris, E., {et~al.} 2021, Astron. J., 162, 148, \dodoi{10.3847/1538-3881/ac1349}

\bibitem[{Whiteford {et~al.}(2023)Whiteford, Glasse, Chubb, Kitzmann, Ray, Phillips, Biller, Palmer, Rice, Waldmann, Changeat, Skaf, Wang, Edwards, \& {Al-Refaie}}]{WhitefordEtAl2023}
Whiteford, N., Glasse, A., Chubb, K.~L., {et~al.} 2023, Mon. Not. R. Astron. Soc., 525, 1375, \dodoi{10.1093/mnras/stad670}

\bibitem[{{Woitke} {et~al.}(2018){Woitke}, {Helling}, {Hunter}, {Millard}, {Turner}, {Worters}, {Blecic}, \& {Stock}}]{WoitkeEtAl2018}
{Woitke}, P., {Helling}, C., {Hunter}, G.~H., {et~al.} 2018, \aap, 614, A1, \dodoi{10.1051/0004-6361/201732193}

\bibitem[{Xuan {et~al.}(2020)Xuan, Bryan, Knutson, Bowler, Morley, \& Benneke}]{xuan_Rotation_2020}
Xuan, J.~W., Bryan, M.~L., Knutson, H.~A., {et~al.} 2020, The Astronomical Journal, 159, 97, \dodoi{10.3847/1538-3881/ab67c4}

\bibitem[{Xuan {et~al.}(2024{\natexlab{a}})Xuan, Hsu, Finnerty, \& Wang}]{XuanEtAl2024b}
Xuan, J.~W., Hsu, C.-c., Finnerty, L., \& Wang, J.~J. 2024{\natexlab{a}}

\bibitem[{Xuan {et~al.}(2022)Xuan, Wang, Ruffio, Knutson, Mawet, Molli{\`e}re, Kolecki, Vigan, Mukherjee, Wallack, Wang, Baker, Bartos, Blake, Bond, Bryan, Calvin, Cetre, Chun, Delorme, Doppmann, Echeverri, Finnerty, Fitzgerald, Horstman, Inglis, Jovanovic, L{\'o}pez, Martin, Morris, Pezzato, Ragland, Ren, Ruane, Sappey, Schofield, Skemer, Venenciano, Wallace, \& Wizinowich}]{XuanEtAl2022}
Xuan, J.~W., Wang, J., Ruffio, J.-B., {et~al.} 2022, Astrophys. J., 937, 54, \dodoi{10.3847/1538-4357/ac8673}

\bibitem[{Xuan {et~al.}(2024{\natexlab{b}})Xuan, Wang, Finnerty, Horstman, Grimm, Peck, Nielsen, Knutson, Mawet, Isaacson, Howard, Liu, Walker, Phillips, Blake, Ruffio, Zhang, Inglis, Wallack, Sanghi, Gonzales, Dai, Baker, Bartos, Bond, Bryan, Calvin, Cetre, Delorme, Doppmann, Echeverri, Fitzgerald, Jovanovic, Liberman, López, Martin, Morris, Pezzato, Ruane, Sappey, Schofield, Skemer, Venenciano, Wallace, Wang, Wizinowich, Xin, Agrawal, Ó, Hsu, \& Phillips}]{XuanEtAl2024a}
Xuan, J.~W., Wang, J., Finnerty, L., {et~al.} 2024{\natexlab{b}}, The Astrophysical Journal, 962, 10, \dodoi{10.3847/1538-4357/ad1243}

\bibitem[{Zahnle \& Marley(2014)}]{ZahnleMarley2014}
Zahnle, K.~J., \& Marley, M.~S. 2014, ApJ, 797, 41, \dodoi{10.1088/0004-637X/797/1/41}

\bibitem[{Zhang {et~al.}(2021{\natexlab{a}})Zhang, Snellen, \& Molli{\`e}re}]{ZhangEtAl2021}
Zhang, Y., Snellen, I. A.~G., \& Molli{\`e}re, P. 2021{\natexlab{a}}, A\&A, 656, A76, \dodoi{10.1051/0004-6361/202141502}

\bibitem[{Zhang {et~al.}(2021{\natexlab{b}})Zhang, Snellen, Bohn, Molli{\`e}re, Ginski, Hoeijmakers, Kenworthy, Mamajek, Meshkat, Reggiani, \& Snik}]{ZhangEtAl2021a}
Zhang, Y., Snellen, I. A.~G., Bohn, A.~J., {et~al.} 2021{\natexlab{b}}, Nature, 595, 370, \dodoi{10.1038/s41586-021-03616-x}

\bibitem[{Zhang {et~al.}(2023)Zhang, Molli{\`e}re, Hawkins, Manea, Fortney, Morley, Skemer, Marley, Bowler, Carter, Franson, Maas, \& Sneden}]{ZhangEtAl2023}
Zhang, Z., Molli{\`e}re, P., Hawkins, K., {et~al.} 2023, Astron. J., 166, 198, \dodoi{10.3847/1538-3881/acf768}

\bibitem[{Zhu {et~al.}(2012)Zhu, Hartmann, Nelson, \& Gammie}]{ZhuEtAl2012}
Zhu, Z., Hartmann, L., Nelson, R.~P., \& Gammie, C.~F. 2012, ApJ, 746, 110, \dodoi{10.1088/0004-637X/746/1/110}

\end{thebibliography}
\bibliographystyle{aasjournal}

\end{document}